\documentclass[aps,reprint,superscriptaddress,amsmath,amssymb]{revtex4-2}

\usepackage{hyperref}
\usepackage[capitalize]{cleveref}
\usepackage{xcolor}
\usepackage{soul}
\usepackage{graphicx}
\usepackage{booktabs}
\usepackage{longtable}
\usepackage{multirow}
\usepackage{pgfplots}

\usepackage{siunitx}[=v2]
\DeclareSIUnit\bit{\textrm{bit}}

\usepackage[version=4]{mhchem}

\usepackage{bibunits}
\defaultbibliographystyle{apsrev4-2}
\defaultbibliography{library}

\DeclareMathOperator{\prob}{\mathrm{P}}
\DeclareMathOperator\erf{erf}

\newcommand{\flabel}[1]{\label{fig:#1}}
\newcommand{\eref}[1]{Eq.~\ref{eqn:#1}}

\newcommand{\elabel}[1]{\label{eqn:#1}}

\newcommand{\mr}[1]{#1}

\makeatletter
\newcommand{\suppresstoc}[1]{{\let\addcontentsline\@gobblethree #1}}
\makeatother

\begin{document}

\begin{bibunit}[apsrev4-2]
\title{ML-PWS: Estimating the Mutual Information Between Experimental
Time Series Using Neural Networks}


\author{Manuel Reinhardt}
\affiliation{AMOLF, Science Park 104, 1098 XG, Amsterdam, The Netherlands}
\author{Gašper Tkačik}
\affiliation{Institute of Science and Technology Austria, 3400
Klosterneuburg, Austria}
\author{Pieter Rein ten Wolde}%
\email{tenwolde@amolf.nl}
\affiliation{AMOLF, Science Park 104, 1098 XG, Amsterdam, The Netherlands}

\renewcommand{\paragraph}[1]{\textit{#1---}}

%\thanks{}

%\textbf\noaffiliation
\date{August 6, 2026}

\begin{abstract}
  \mr{The ability to quantify information transmission is crucial for
    the analysis and design of both natural and engineered systems.
    For systems driven by time-varying signals, the fundamental
    measure is the information transmission rate. However, due to the
    high dimensionality of signal trajectory space, this rate cannot
    be obtained directly from time-series data without
    approximations. Path Weight Sampling (PWS) is a computational
    technique that enables the exact calculation of the information
    rate for any stochastic model, raising the question of how this
    rate can be determined from time-series data in the absence of a
    prior model.
    Here, we present a method that combines machine learning (ML)
    with PWS: a generative model is learned from time-series data,
    to which PWS is applied to yield a rigorous lower bound on the
  information rate.} We demonstrate the accuracy of this technique,
  called ML-PWS, by comparing its results on synthetic time-series
  data generated from several non-linear models against ground-truth results
  obtained by applying PWS directly to the same models. We illustrate
  the utility of ML-PWS by applying it to neuronal time-series data.
\end{abstract}

%\keywords{}

\maketitle

The canonical measure for quantifying information transmission via
time-varying signals is the information transmission rate
\cite{1948.Shannon, 2009.Tostevin}. It
quantifies the speed at which distinct messages are transmitted
through the system, accounting for the correlations
in the input and output signals.
The information rate has been used to quantify biochemical signaling
performance
\cite{2021.Mattingly,2023.Reinhardt,2023.Hahn,2023.Moor,2025Moor},
to perform model reduction \cite{2023.Schmitt},
to detect the causality of interactions
\cite{2007.Frenzel,2007.Hlavackova-Schindler},
% to test for nonlinearities in time series \cite{1995.Palus},
to assess dependencies between
stock prices over time
\cite{2002.Marschinski,2013.Dimpfl,2016.Dimpfl}, or to quantify
information exchange
between different brain regions \cite{2011.Rad,2012.So}. In the
absence of feedback the rate also equals the multi-step transfer entropy
\cite{1990.Massey,2000.Schreiber}.

Yet, computing the path mutual information between input and output
trajectories is notoriously difficult because trajectories are
high-dimensional. Conventional methods rely on non-parametric
estimates of the joint probability
distribution of input and output, e.g., histograms or kernel
density methods \cite{1986.Fraser,1995.Moon}. These are,
however, infeasible for high-dimensional data
\cite{2003.Paninski,2019.Cepeda-Humerez}.  More advanced
non-parametric estimators such as the k-nearest-neighbor (KNN)
estimator \cite{2004.Kraskov} perform better, but still suffer from
uncontrolled biases as the dimensionality increases
\cite{2014.Gao,2019.Cepeda-Humerez,das2025exact-3b4}.
Consequently, the rate is often computed using approximate methods,
such as the Gaussian framework  \cite{2009.Tostevin,2010.Tostevin},
moment-closure approximations \cite{2019.Duso,2023.Moor}, or
techniques limited to specific systems
\cite{Sinzger.2020,2023.Sinzger,Gehri.2024dmt}.
Machine-learning schemes have also been introduced, such as
decoding-based approaches
\cite{quiroga2009extracting-7fd,2018Granados}, neural variational
estimators
\cite{2004.Barber,2010.Nguyen,2018Belghazi,2019.Poole,song2020understanding-e06},
noise-contrastive estimators (InfoNCE) \cite{2018.Oord} and the
Difference-of-Entropies (DoE) estimator \cite{2018.McAllester}.

We recently presented Path Weight Sampling (PWS)
\cite{2023.Reinhardt}, a computational technique
for exactly calculating the information transmission rate of
stochastic models described by master equations and Langevin equations. PWS uses the model
to evaluate the
path likelihood, i.e. the
conditional probability of an individual output trajectory for a
given input trajectory, and then averages it via Monte
Carlo sampling in trajectory space to obtain the path mutual
information. While the scheme involves no approximations, it
requires a stochastic model to evaluate
the path likelihood and therefore cannot be applied directly
to experimental time-series data.

In this manuscript, we show how PWS can be combined with machine
learning to obtain the information rate directly from experimental
time-series data. The idea of ML-PWS is to first learn a
generative model of the time-series data, which yields the path
likelihood, and then employ PWS to obtain the mutual
information.
%We show that when this path likelihood is then averaged to obtain
% the information transmission rate, we obtain a lower bound of the
% information rate.
In contrast to other techniques based on generative models, such as
DoE  \cite{2018.McAllester}, ML-PWS provides a rigorous lower bound on
the information rate.
We train
autoregressive sequence prediction models, as commonly used e.g. in speech
synthesis \cite{2016.Oord} or text generation \cite{2011.Sutskever},
but the principal idea is more generic.
We demonstrate, for several
examples, the power of ML-PWS by applying it to synthetic time series
data generated from a nonlinear model, and by comparing the
rate thus obtained against the ground-truth result obtained by
applying PWS directly to
this model. We further illustrate the utility of ML-PWS  by applying
it to neuronal time-series data \cite{Marre.2015}.

\mr{
  \paragraph{Setup}%
  We consider a system stimulated by a time-varying input
  $s(t)$ that responds with an output $x(t)$.
  We assume input and output are recorded as trajectories
  $s_{1:n}=(s(\delta t),\ldots,s(n\delta t))^\intercal$ and
  $x_{1:n}=(x(\delta t),\ldots,x(n\delta t))^\intercal$, sampled at
  timestep $\delta t$.
  We further assume the input is controlled or
  characterized by the experimenter, so its distribution
  $\prob(s_{1:n})$ is known.
  From repeated experiments, one obtains a dataset of $N$ pairs of
  discretized trajectories $(s^k_{1:n}, x^k_{1:n})$, $k = 1, \ldots, N$.
  The quantity of interest is the information transmission rate
  $R(\mathcal{S}; \mathcal{X})$, defined as the asymptotic rate at
  which the path
  mutual information between input and output trajectories grows with
  trajectory duration $T = n\,\delta t$,
  \begin{equation}
    R(\mathcal{S};\mathcal{X}) = \lim_{n\to\infty}
    \frac{1}{n\,\delta t}\,I(S_{1:n}; X_{1:n})\,.
  \end{equation}
  ML-PWS is a scheme to compute a lower bound for
  $R(\mathcal{S};\mathcal{X})$ from the
  experimental dataset.%
}

\mr{
  \paragraph{Monte-Carlo estimate of the mutual information}%
  To compute the mutual information via Monte Carlo averages we write
  it as a difference of entropies,
  \begin{equation}
    I(S_{1:n}; X_{1:n}) = H(X_{1:n}) - H(X_{1:n}|S_{1:n}),
    \label{eq:mi_entropy_diff}
  \end{equation}
  with the marginal entropy $H(X_{1:n}) =
  -\langle\ln\prob(x_{1:n})\rangle$ and the conditional entropy
  $H(X_{1:n}|S_{1:n}) = -\langle\ln\prob(x_{1:n}|s_{1:n})\rangle$.
  The PWS scheme is based on the insight that both
  entropies can be computed as Monte Carlo
  averages in path space \cite{2023.Reinhardt}.
  Yet, these averages require two key quantities: the true path
  likelihood $\prob(x_{1:n}|s_{1:n})$ and the true marginal
  $\prob(x_{1:n}) = \int ds_{1:n} \prob(s_{1:n})\prob(x_{1:n}|s_{1:n})$.
  In the absence of a mechanistic model, the path likelihood must be
  estimated from experimental data.
}

\mr{
  \paragraph{Learning the path likelihood}%
  To estimate the path likelihood we use an
  autoregressive model, which factorizes it as
  $\prob(x_{1:n}|s_{1:n})=\prod_i \prob(x_i|s_{1:i},x_{1:i-1})$. This
  expression is exact: each $x_i$ depends on the
  full history of the input $s_{1:i}$ and output $x_{1:i-1}$.
  To make further progress, we use an
  autoregressive Gaussian model, approximating the true path likelihood as
  \begin{align}
    Q(x_{1:n} \mid s_{1:n}) = \prod^n_{i=1}
    \frac{1}{\sqrt{2\pi} \sigma_i} \exp\left(-\frac{(x_i -
    \mu_i)^2}{2\sigma_i^2}\right)
    \label{eq:autoregressive_conditional_density}.
  \end{align}
  We use a deep neural network to predict the mean $\mu_i$ and
  standard deviation $\sigma_i$, which depend on the full trajectory
  histories $s_{1:i}$ and $x_{1:i-1}$. Notably, although each factor
  in \cref{eq:autoregressive_conditional_density} $q(x_i |s_{1:i},
  x_{1:i-1}) = \mathcal{N}(x_i; \mu_i,\sigma_i)$ is Gaussian,
  the distribution $Q(x_{1:n}\mid s_{1:n})$ is not, because $\mu_i$
  and $\sigma_i$ depend nonlinearly on these histories.
  More expressive generative models, e.g. normalizing flows, can
  replace the Gaussian conditional when $\prob(x_i\mid
  s_{1:i},x_{1:i-1})$ is itself non-Gaussian.
  The resulting model is trained via gradient descent by minimising
  the negative log-likelihood $\mathcal{L} = -\sum^N_{k=1} \ln
  Q(x^k_{1:n} \mid s^k_{1:n})$ over the dataset.
}

\mr{
  \paragraph{Computing the marginal path likelihood}%
  We define the marginal path likelihood of our model as
  \begin{equation}
    Q(x_{1:n}) \equiv
    \int ds_{1:n}\,\prob(s_{1:n})Q(x_{1:n}|s_{1:n})\,.
    \elabel{marginal}
  \end{equation}
  Since we assume the input distribution $\prob(s_{1:n})$ is
  known, we can compute this integral via a Monte Carlo scheme from
  the path likelihood without requiring a separate model for the
  marginal probability.
  Computing this integral efficiently in path space is a challenge
  addressed by our earlier work \cite{2023.Reinhardt},
  in which we showed that sequential Monte Carlo and free-energy
  estimation techniques can significantly enhance the computational
  efficiency (see Sec.~\ref{sec:SM-marginalisation} of
  Supplemental Material, SM).
}

\mr{
  \paragraph{A lower bound on the mutual information}%
  From $Q(x_{1:n}|s_{1:n})$ and $Q(x_{1:n})$, the mutual
  information is estimated as $\hat{I}_Q = \hat{H}_Q(X) - \hat{H}_Q(X|S)$ with
  \begin{align}
    \hat{H}_Q(X|S) &= -\frac{1}{N}\sum^N_{i=1}\ln Q(x^i_{1:n}|s^i_{1:n}), \\
    \hat{H}_Q(X)   &= -\frac{1}{N}\sum^N_{i=1}\ln Q(x^i_{1:n}).
    \elabel{MI_estimate}
  \end{align}
  These quantities are estimators for the cross-entropy between the
  true underlying distribution of the data, $P$, and the neural
  network approximation, $Q$.
  The cross-entropy over-estimates the true entropy unless $Q=P$: the
  bias of $\hat{H}_Q(X|S)$ is
  $\mathbb{E}_{\prob(s)}\{D_\mathrm{KL}[\prob(x|s)\Vert Q(x|s)]\} \geq 0$
  and the bias of $\hat{H}_Q(X)$ is $D_\mathrm{KL}[\prob(x)\Vert Q(x)]
  \geq 0$.
  The key result, which we prove in Sec.~\ref{sec:SM-mi-lower-bound}
  of the SM, is that the
  latter never exceeds the former
  \begin{equation}
    D_\mathrm{KL}[\prob(x)\Vert Q(x)]
    \leq
    \mathbb{E}_{\prob(s)}\bigl\{D_\mathrm{KL}[\prob(x|s)\Vert Q(x|s)]\bigr\},
    \elabel{kl_ineq}
  \end{equation}
  so the bias in the conditional entropy (second term in
  \cref{eq:mi_entropy_diff}) dominates over that in the marginal, and
  we obtain a lower bound for the mutual information:
  \begin{equation}
    \hat{I}_Q(S_{1:n}; X_{1:n}) \leq I(S_{1:n}; X_{1:n}).
    \elabel{lower_bound}
  \end{equation}
}

\mr{
  The inequality \eref{kl_ineq} holds because $Q(x_{1:n})$ is the
  exact marginal of $Q(x_{1:n}|s_{1:n})$ under $\prob(s_{1:n})$.
  If instead the marginal were approximated by a separately trained
  network no bound can be derived.
  The bound tightens monotonically as the generative model improves,
  and matches the true mutual information if and only if
  $Q(x_{1:n}|s_{1:n}) = \prob(x_{1:n}|s_{1:n})$.
  To obtain this rigorous lower bound, ML-PWS must be evaluated on
  the {\em measured} responses, which we refer to as {\em data mode}.
}

\begin{figure*}
  \centering
  \includegraphics[width=\textwidth]{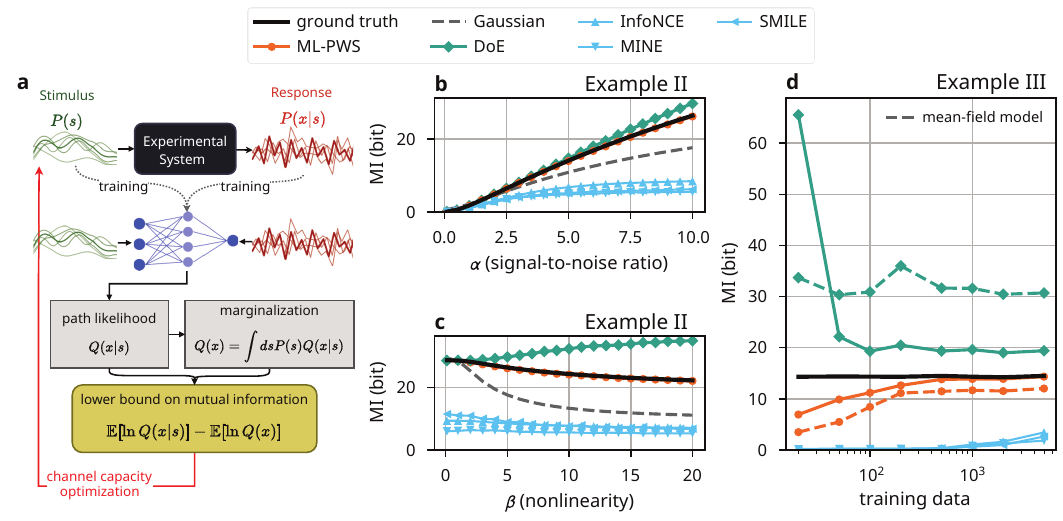}
  \caption{
    \mr{Comparison of ML-PWS against the ground truth and alternative
      mutual information estimators for two of the three model systems.
      Ground truth (black) is computed by applying PWS directly to the
      known model dynamics. (a) Overview of the ML-PWS method.
      (b,c) Example~II: MI $I(S_{1:30};X_{1:30})$ as a function of the
      signal-to-noise ratio $\alpha$ at fixed nonlinearity $\beta =
      5.0$ (b), and as a function of $\beta$ at fixed $\alpha = 10.0$ (c).
      (d) Example~III: MI as a function of training set size
      $N_\text{train}$ for a system with a
      two-dimensional, correlated output.
      Dashed lines: ML-PWS and DoE under a
      deliberately misspecified mean-field model.
  }}
  \flabel{benchmarks}
\end{figure*}

\mr{
\paragraph{Estimating channel capacity} The learned model can also be
  used to generate {\em new} responses to inputs drawn from a
  different distribution. Although this {\em generative mode} does not
  itself yield a strict bound, it enables the estimation of the
  channel capacity.
  In systems without feedback, $\prob(x_{1:n}|s_{1:n})$ is a property
  of the system, independent of the stimulus statistics. A generative
  model $Q(x_{1:n}|s_{1:n})$ that accurately captures this
  input--output mapping can therefore be used to predict the
  mutual information for input distributions other than the one used
  in the original experiment, without retraining.
  ML-PWS can thus search over input distributions to maximise
  the information rate---a channel-capacity prediction.
  This prediction can then be tested experimentally:
  repeating the experiment with the proposed stimulus distribution
  yields a new dataset from which a strict new lower bound can be
  computed (\eref{lower_bound}).
  Each round of experiments can thus raise the established lower bound on the
  channel capacity, cf. \cref{fig:benchmarks}(a). %
}

\paragraph{Benchmarks}%
We validate ML-PWS on three synthetic systems whose ground-truth mutual
information
$I(S_{1:n};X_{1:n})$ follows from applying PWS directly to the
stochastic models.
\mr{
  We compare our results
  against the Gaussian approximation~\cite{2010.Tostevin},
  DoE~\cite{2018.McAllester}, SMILE~\cite{song2020understanding-e06},
  MINE~\cite{2018Belghazi}, and InfoNCE~\cite{2018.Oord}.
  The variational estimators (InfoNCE, MINE, SMILE), as well as
  ML-PWS are lower bounds by
  construction, and the Gaussian approximation yields a lower bound
  only when the input statistics are Gaussian \cite{2001.Mitra};
  DoE is the only estimator we tested that does not yield a bound.
  To allow for a fair comparison, we used the same NN architectures
  for the different estimators whenever possible (see SM).
  Example~I, a simpler variant of Example~II with an AR(3) input, is
  presented in Sec.~\ref{sec:SM-example1} of the SM.
}

\mr{
  \paragraph{Example II: minimal nonlinear channel}%
  The input $S_t$ is a sum of two AR(1) processes (see SM); the scalar
  output follows $X_t=\alpha\,f_\beta(S_t)+\rho\,X_{t-1}+\vartheta\,\eta_t$,
  with $f_\beta$ a saturating sigmoid.
  Each trajectory in the training set ($N_\text{train}=1000$) comprises $n=30$
  time steps and spans about one
  correlation time of the slowest input mode.
  The parameters $\alpha$ and $\beta$ cleanly separate two distinct
  aspects of the output model: $\alpha$ sets the signal-to-noise ratio (SNR),
  while $\beta$ controls the degree of nonlinearity. Varying each
  independently [\cref{fig:benchmarks}(b,c)] exposes the regimes in which
the various estimators succeed or fail.}

\mr{We find that ML-PWS significantly outperforms the other lower bounds.
  Since we consider relatively long trajectories we are in the
  high-MI regime, where the variational bounds InfoNCE,
  MINE and SMILE are known to be limited by sample size~\cite{2018.Oord,2019.Poole,song2020understanding-e06} and
  therefore severely
  underestimate the MI.
  Generally, the Gaussian approximation is only exact
  for linear systems but because the input here is
  Gaussian it still yields a strict
  lower bound for a nonlinear system~\cite{2001.Mitra}.
  The Gaussian bound becomes increasingly loose
  as the nonlinearity $\beta$ increases, and for a strongly nonlinear
  channel at large SNR it lies well below the ML-PWS bound. ML-PWS
  does not have the limitations of the other methods and stays
  indistinguishable from the
ground truth throughout.}

\begin{figure*}
  \centering
  \includegraphics[width=\textwidth]{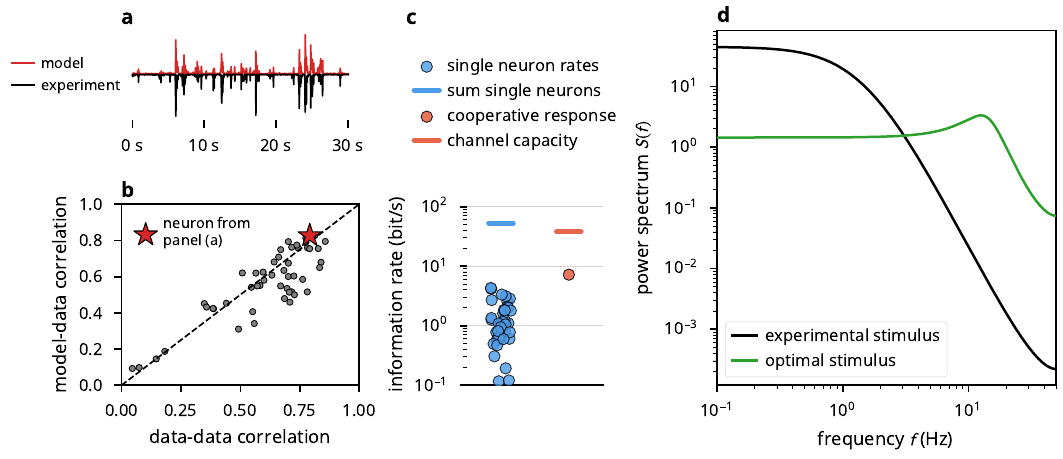}
  \caption{\mr{Model validation, information rate and channel capacity
    for a population of neurons. (a) Comparison of the response to a
    given time-varying stimulus of a single-neuron model (red) against
    an experimental recording (black). (b)  Model-data correlation of
    the collective model
    (correlation coefficient between the response predicted by the
    collective model and the recorded response) vs.
    data-data correlation (correlation coefficient between responses
    measured in two independent halves of the repeated trials) for all
    50 neurons (see Sec.~\ref{sec:SM-ganglion-validation} of the SM).
    Each point is one neuron; the star marks the
    neuron shown in (a). (c) Information rates (bit/s) estimated using
    ML-PWS. Blue dots: individual single-neuron models; blue bar: their
    sum. Orange dot and bar: the joint model and the channel capacity
    (obtained by optimizing stimulus statistics),
    respectively. (d) Power spectrum of the capacity-maximizing input
  stimulus (green) compared against the experimental input spectrum (black).}}
  \flabel{NR}
\end{figure*}

\mr{DoE, by contrast, \emph{overestimates} the true MI at large~$\beta$.
  The origin is an asymmetry between
  the conditional and the marginal. Here the one-step conditional
  $\prob(x_i\mid s_{1:i},x_{1:i-1})$ is Gaussian, but the one-step
  marginal $\prob(x_i\mid x_{1:i-1})$ is Gaussian only for small $\beta$,
  so DoE's separately trained Gaussian marginal network mis-estimates
  the marginal entropy $H(X)$.
  %   ML-PWS instead obtains the marginal entropy by
  %   exact Monte Carlo marginalization of the learned conditional
  %   probability density and thereby
  % avoids this failure.
}

\mr{
  \paragraph{Example III: variance-modulated two-channel output}%
  In this system the signal is encoded entirely in the output
  \emph{(co)variance}. A scalar AR(1) signal $S_t$ modulates the common
  amplitude of a shared latent AR(1) factor $Z_t$ that drives both
  components of a two-dimensional output
  $\boldsymbol{X}_t\equiv(X^1_t,X^2_t)$ (see SM). Because the signal
  enters only through the (co)variance, it leaves the output mean
  $\langle \boldsymbol{X}_t \rangle = 0$, irrespective of the signal.
  Hence, $S$ and $\boldsymbol{X}$ are linearly uncorrelated, and the
  Gaussian approximation cannot be used. The per-step conditional
  $\prob(\boldsymbol{x}_t | s_{1:t},\boldsymbol{x}_{1:t-1})$
  is exactly
  Gaussian and analytically tractable, whereas the marginal
  $\prob(\boldsymbol{x}_t|\boldsymbol{x}_{1:t-1})$ is a heavy-tailed
  mixture of Gaussians that is difficult to learn from data.
  Since the conditional probability is tractable but the marginal is not,
  we expected ML-PWS to perform better than DoE in this benchmark
  since the latter requires
  learning the marginal distribution from data.
}

\mr{
  Varying the training-set size $N_\text{train}$
  [\cref{fig:benchmarks}(d)] yields two
  conclusions. First, ML-PWS significantly
  \emph{underestimates} the MI when data are scarce as any lower bound
  must, but converges to the ground truth by $N_\text{train}\approx 10^3$;
  DoE instead plateaus at an overestimate.
  The variational estimators detect near-zero MI. Second, the ML-PWS
  lower bound is robust to misspecification of the output covariance:
  while the model that allows for cross-correlations between the two
  output components reaches the ground-truth result (solid line), a
  deliberately misspecified mean-field (diagonal covariance) model
  that cannot capture these correlations still yields a valid lower
  bound (dashed line).
  DoE does not yield a bound in either case, performing
  significantly worse with the misspecified model.
  The model definition and training hyperparameters are given in the SM.
}

\paragraph{Application to neural data}To demonstrate the utility of
ML-PWS, we apply it to neuronal time series data \cite{Marre.2015} (\cref{fig:NR}).
The data was recorded from retinal ganglion cells of a salamander,
responding to a
dark horizontal bar, displayed on a screen, moving stochastically in
a vertical direction \cite{Marre.2015}. \mr{We used spike trains of
50 neurons.}
The dynamics of the stimulus, the moving bar, is that of an
underdamped particle in a harmonic well, which means that the
input distribution, $\prob(s_{1:n})$, is known.
The time series of the output was discretized at a
resolution of $\Delta t = 10\,{\rm ms}$.
\mr{
  The non-repeating parts of the recordings were used for model
  training and the ML-PWS estimate, while the stimulus-repeat trials
  were reserved for model validation
  (see Sec.~\ref{sec:SM-ganglion} of the SM).
}
We trained a neural network for each cell separately, yielding a
generative model for the path likelihood
$\prob(x^{(i)}_{1:n}|s_{1:n})$ for each cell $i$, from which we then
computed the path mutual information  $I(S_{1:n};X^{(i)}_{1:n})$ for
each individual cell. This model assumes that, for a given stimulus,
the number of spikes at a given time is Poisson distributed. In
addition, we trained a neural network for the collective response of
the population of cells, yielding the path likelihood
$\prob({\boldsymbol x}_{1:n}|s_{1:n})$ for the population of cells, with
${\boldsymbol x}_{1:n}= (x^{(1)}_{1:n},\dots,x^{(50)}_{1:n})$; this
yields the mutual information $I(S_{1:n};{\boldsymbol X}_{1:n})$
between the input $S_{1:n}$ and the collective response ${\boldsymbol
X}_{1:n}$ of the population of cells. \mr{From these mutual
informations we computed the corresponding information rates (see SM).}

\mr{\Cref{fig:NR}(a) shows that the generative model can accurately
  produce spike trains that match the experimental time series data
  for an individual neuron.
  \Cref{fig:NR}(b) quantifies, for each neuron, the correlation of the
  collective model output with the experimental validation data and
  compares it against
  the correlation \emph{within} independent validation trials.
  The latter is the noise ceiling: the trial-to-trial variability of
  the neurons limits how well \emph{any} model can predict a single
  response.
  The points scatter around the diagonal, with the model-data
  correlation reaching on average $95\%$ of the data-data correlation,
  which shows that the model predicts the neural responses nearly as
well as the data predict themselves, across the 50-neuron dataset.}
The information rate between the input and the collective response
(\cref{fig:NR}(c), orange dot) is significantly lower than the sum of
the mutual information of the individual cells: correlations between
the responses of the individual cells lead to redundant coding,
lowering the information that is encoded in the collective response
\cite{Marre.2015}.

\mr{
  We also computed the channel capacity for the collective response
  of the 50 neurons. To do so, we optimized the two
  parameters of the input subject to a power constraint.
  \Cref{fig:NR} shows that the optimal input distribution (d) yields
  a drastically higher information rate than that used experimentally (c).
}

\mr{
  \paragraph{Discussion}We demonstrated how we can combine
  autoregressive sequence-prediction models, trained on time-series
  data, with the PWS Monte Carlo estimator to obtain a tight lower
  bound on the information rate. Benchmarking ML-PWS against the
  Gaussian approximation and four machine-learning estimators (DoE,
  InfoNCE, MINE, SMILE) on three synthetic systems, we found that ML-PWS
  is the most accurate estimator: the variational bounds saturate well
  below the true MI, the Gaussian approximation becomes worse with
  increasing nonlinearity of the output, and DoE, which trains two
  neural networks to separately estimate the conditional and marginal
  distributions, systematically overestimates the true MI in our tests.
  Because ML-PWS instead obtains the marginal probability by exact
  Monte Carlo marginalization of the learned conditional probability,
  it yields a reliable lower bound even when data is scarce or the
  generative model is misspecified (\cref{fig:benchmarks}d). Finally,
  we note that ML can be combined with TE-PWS~\cite{das2025exact-3b4}
  to estimate transfer entropies and causation entropies directly
  from time series data.
}

\paragraph{Acknowledgments}This work is part of the Dutch Research
Council (NWO) and was performed at the research institute AMOLF.
This project has received funding from the European Research Council
(ERC) under the European Union’s Horizon 2020 research and innovation
program (grant agreement No.~885065),
and was financially supported by NWO through the “Building a
Synthetic Cell (BaSyC)” Gravitation grant (024.003.019). GT
acknowledges the support of WWTF grant LS23-026 ``Understanding
pancreas biology with AI/ML''.

\paragraph{Code Availability}The code used to generate the results of
this study is openly available \cite{manuel-rhdt_ml-pws_2025}.

\suppresstoc{\putbib[library]}
\end{bibunit}

% ============================================================
% SUPPLEMENTAL MATERIAL
% ============================================================

\clearpage
\onecolumngrid

\begin{bibunit}[apsrev4-2]
\suppresstoc{\section*{Supplemental Material}}

% Standard APS convention for supplemental material: prefix all
% section, equation, figure, and table numbers with "S" so that they
% cannot be confused with the corresponding numbering in the main text.
\renewcommand{\thesection}{S\arabic{section}}
\renewcommand{\theequation}{S\arabic{equation}}
\setcounter{equation}{0}
\renewcommand{\thefigure}{S\arabic{figure}}
\renewcommand{\thetable}{S\arabic{table}}

\makeatletter
\let\l@subsubsection\@gobbletwo
\let\l@paragraph\@gobbletwo
\makeatother
\tableofcontents

\newpage

This Supplemental Material provides derivations, additional analyses,
and implementation details supporting the results presented in the
main text.

\section*{Notation}

Throughout this document, expectations are denoted by
$\mathbb{E}_{P(x)}[f(x)] := \int dx\,P(x)\,f(x)$, with sums replacing
integrals for discrete variables. The KL-divergence between the
distributions $P(x)$ and $Q(x)$ is defined as
\begin{equation}
  D_\text{KL}\big[P(x)\,\Vert\,Q(x)\big] \equiv
  \mathbb{E}_{P(x)}\left[\ln\frac{P(x)}{Q(x)}\right]\,.
\end{equation}

We consider a system with stochastic input $S$ and output $X$; our
approach is generic and applies to arbitrary high-dimensional
distributions. A central application is the case where $S$ and $X$
are discrete-time stochastic trajectories, written as
$x_{1:n} = (x_1,\ldots,x_n)$, and analogously for $s_{1:n}$.
We denote by $P(s)$, $P(x|s)$,
and $P(s,x)$ the marginal, conditional, and joint distributions of the
data, respectively. The marginal and conditional entropies are defined as
\begin{align}
  H(X) &\equiv \mathbb{E}_{P(x)}[-\ln P(x)]\,, \\
  H(X\mid S) &\equiv \mathbb{E}_{P(s,x)}[-\ln P(x|s)]\,,
\end{align}
and the mutual information between $S$ and $X$ is
\begin{equation}
  I(S;X) \equiv H(X) - H(X\mid S)\,.
\end{equation}
For stationary processes $\mathcal{S}, \mathcal{X}$, the mutual
information rate (or short: information rate) is defined as
\begin{equation}
  R(\mathcal{S}, \mathcal{X}) \equiv \lim_{n\to\infty}
  \frac{1}{n\,\Delta t}\,I(S_{1:n}; X_{1:n})\,,
\end{equation}
where $\Delta t$ is the discrete time step.

\section{Proof of the Mutual Information Lower Bound}
\label{sec:SM-mi-lower-bound}

ML-PWS provides a lower bound for the mutual information (MI)
$I(S;X) = H(X) - H(X\mid S)$ by replacing each entropy with a
\emph{cross-entropy} with respect to a learned model $Q(x|s)$.
For the two probability densities $P(x|s)$ and $Q(x|s)$, the
conditional cross-entropy decomposes as
\begin{equation}
  \label{eq:SM-cent1}
  H_Q(X\mid S) \equiv \mathbb{E}_{P(s,x)}[-\ln Q(x|s)]
  = H(X\mid S) + \mathbb{E}_{P(s)}\big\{
  D_\text{KL}\big[P(x|s)\,\Vert\,Q(x|s)\big]\big\}
\end{equation}
and similarly for the marginal entropy
\begin{equation}
  \label{eq:SM-cent2} H_Q(X) \equiv
  \mathbb{E}_{P(x)}[-\ln Q(x)] = H(X) +
  D_\text{KL}\big[P(x)\,\Vert\,Q(x)\big]\,.
\end{equation}
% where $Q(x)$ is defined as $\mathbb{E}_{P(s)}[Q(x|s)]$ \footnote{%
%   In any practical implementation $Q(x)$ is replaced by a finite-$M$
%   Monte Carlo estimate $\hat{Q}(x) = M^{-1}\sum^M_{i=1} Q(x|s_i)$
%   with $s_i\sim\prob(s)$. By Jensen's inequality applied to the
%   concave logarithm, $\langle\ln \hat{Q}(x)\rangle \leq \ln Q(x)$,
%   with equality only in the absence of Monte Carlo variance. The
%   finite-$M$ marginal-entropy estimator
%   $\hat{H}^{\prime\prime}_\text{PWS}(X) =
%   \mathbb{E}_{P(s,x)}[-\ln\hat{Q}(x)]$ therefore satisfies
%   $\hat{H}^{\prime\prime}_\text{PWS}(X)\geq
%   \hat{H}^\prime_\text{PWS}(X)$, while the conditional-entropy
%   estimator carries no analogous Monte Carlo bias. As a result the
%   finite-$M$ MI estimator is positively biased relative to its
%   $M\to\infty$ value $\hat{I}^\prime$ and is not itself a guaranteed
%   lower bound on $I(S;X)$; it converges to one as $M\to\infty$.
%   ML-PWS is, in this sense, a consistent Monte Carlo estimator for a
%   mutual-information lower bound. The same caveat applies to every
%   variational MI estimator that approximates a marginal density by
%   Monte Carlo, including the contrastive estimators discussed in the
%   next section. The finite-$M$ bias is quantified empirically in
%   \cref{sec:SM-bias}.
% }.
Due to the non-negativity of the KL-divergence, both
cross-entropies $H_Q(X | S)$ and $H_Q(X)$
over-estimate the true entropies $H(X|S)$ and $H(X)$.
Nevertheless, we will show that the difference
\begin{equation}
  I_Q(S;X) = H_Q(X) - H_Q(X | S)
\end{equation}
between these cross-entropies
results in a lower bound for the mutual information.

To this end, define the joint distribution
$Q(s,x):=P(s)Q(x|s)$. This distribution is well-defined and has the
marginal $Q(x)=\int ds\,Q(s,x)=\mathbb{E}_{P(s)}[Q(x|s)]$. By analogy
to Bayes' theorem
we further define \footnote{%
  This construction relies on $Q(x)$ being the exact marginal of
  $Q(x|s)$. If $Q(x)$ were replaced by a separately trained marginal
  model $Q^\prime(x)$, as in \cite{2020.McAllester}, then
  $P(s)Q(x|s)/Q^\prime(x)$ would no longer be a normalized
  distribution. With the Monte Carlo marginal $\hat{Q}(x)$ used in
  ML-PWS, $Q(s|x)$ is normalized only in the limit $M\to\infty$.
}
\begin{equation}
  Q(s|x) := \frac{Q(s,x)}{Q(x)} = \frac{P(s)Q(x|s)}{Q(x)}\,.
  \label{eq:SM-Qsx}
\end{equation}

Combining \cref{eq:SM-cent1,eq:SM-cent2} and then using
definition \eqref{eq:SM-Qsx}, we derive
\begin{equation}
  I_Q(S;X)
  = \mathbb{E}_{P(s,x)}\!\left[\ln \frac{Q(x|s)}{Q(x)}\right]
  = \mathbb{E}_{P(s,x)}\!\left[\ln \frac{Q(s|x)}{P(s)}\right]\,.
  \label{eq:SM-Ihat}
\end{equation}
This expression has the same form as
$I(S;X)=\mathbb{E}_{P(s,x)}[\ln P(s|x)/P(s)]$ but with $Q(s|x)$ in
place of $P(s|x)$. Subtracting the two, we find
\begin{equation}
  I(S;X) - I_Q(S;X)
  = \mathbb{E}_{P(s,x)}\!\left[\ln \frac{P(s|x)}{Q(s|x)}\right]
  = \mathbb{E}_{P(x)}\!\left\{D_\text{KL}[P(s|x) \mathrel{\Vert} Q(s|x)]\right\}
  \geq 0\,,
  \label{eq:SM-gap}
\end{equation}
by the non-negativity of the KL-divergence. Hence
$I_Q(S;X) \leq I(S;X)$, with equality if and only if
$Q(s|x)=P(s|x)$.

Note that the lower-bound in \cref{eq:SM-gap} holds for
\emph{any} model $Q(x|s)$: the quality of the trained network
controls the tightness of the bound. ML-PWS is a Monte Carlo
estimator for that lower bound based on a neural network model
$Q(x|s)$ trained using a negative log-likelihood loss on samples from $P(s,x)$.
The KL-divergence in \cref{eq:SM-gap} therefore quantifies the model
mismatch: the gap between $I(S;X)$ and $I_Q(S;X)$ is precisely the
average discrepancy between the exact posterior $P(s|x)$ and the
posterior $Q(s|x)$ implied by the model.

A practical consequence is that the estimated
bound $I_Q(S;X)$ can be \emph{negative}, even though the true mutual
information satisfies $I(S;X)\geq0$. Indeed, a
lower bound on a non-negative quantity need not itself be
non-negative. From \cref{eq:SM-gap}, we see that the bound becomes
negative when the model mismatch exceeds the true
information. This occurs most commonly at short trajectory durations
$T$, where $I(S;X)$ is small.
A negative value therefore signals a loose (uninformative) bound at
that duration rather than a failure of the method, and the estimate
becomes informative once $T$ is large enough.

\section{Estimating the Mutual Information of the Model from Generative
Samples}\label{sec:SM-model_mi}

In addition to being able to compute a lower bound on the mutual
information, ML-PWS can also be
used to compute the MI of the generative model itself, by sampling
from the model rather than the real system.
Specifically, whether ML-PWS returns a true lower bound is
determined solely by
the distribution from which the sample pairs $(s,x)$ are drawn.
If the responses $x$ are
taken from the measured data, thus produced by the real system,
i.e.\ $x \sim P(x|s)$, the estimator converges to
$I_Q(S;X) \leq I(S;X)$, as shown in the previous section.

If instead the responses are \emph{generated by the model},
$x \sim Q(x|s)$, the sample pairs are distributed according to
$Q(s,x) = P'(s)\,Q(x|s)$, where $P'(s)$ is any input ensemble,
possibly distinct
from $P(s)$; the estimator then converges to
\begin{equation}
  \mathbb{E}_{Q(s,x)}\!\left[\ln \frac{Q(x|s)}{Q(x)}\right] = I_{Q}'(S;X)\,,
  \label{eq:SM-model_mi}
\end{equation}
which is the mutual information \emph{of the generative model} under the
input statistics $P'(s)$. Importantly, this quantity is not a lower bound
for the true mutual information $I(S;X)$.
Yet, it quantifies the MI exactly \emph{for the model} since
the likelihood $Q(x|s)$ and the input distribution $P'(s)$ are known, so
the only remaining errors are the Monte Carlo variance and the bias incurred
when marginalizing $Q(x)$, which we quantify in \cref{sec:SM-bias}.

This mode of operation is useful for using the trained network
with input statistics that are not accessible experimentally, for instance,
for computing the channel capacity.
It must, however, be interpreted with care:
\cref{eq:SM-model_mi} characterizes the fitted model rather than the real
system, and it carries no guarantee with respect to $I(S;X)$ unless the
model is accurate.
Only when $x$ is sampled from the real, measured output data does the
estimate have the lower-bound property of \cref{eq:SM-gap}.

\section{ML-PWS as a Special Case of the Donsker--Varadhan Bound}
\label{sec:SM-dv-bound}

This section serves two purposes. First, we provide an
alternative proof that ML-PWS estimates a lower bound on the mutual
information, complementing the direct argument of the previous
section. Second, we clarify how ML-PWS is related to variational
MI estimators such as MINE and InfoNCE that estimate the
Donsker--Varadhan (DV) lower bound for the mutual information.
We show that these methods differ from ML-PWS in how they
compute that bound.

% \citet{song2020understanding-e06} show that variational MI estimation
% can be viewed as a constrained optimization over density ratios. The
% density ratio relevant for the mutual information is
% \begin{equation}
%   r(s,x) = \frac{P(s,x)}{P(s)\,P(x)} = \frac{P(x|s)}{P(x)}\,.
% \end{equation}
% The mutual information can then be expressed as
% \begin{equation}
%   I(S;X) = \mathbb{E}_{P(s,x)}[\log r(s,x)]\,.
% \end{equation}
% While the exact density ratio $r(s,x)$ is generally unknown, the
% mutual information can be lower-bounded using approximations of this
% ratio. The mathematical framework for this follows from the
% Donsker--Varadhan inequality.

\subsection{The Donsker--Varadhan Lower Bound}

The variational mutual information estimators, like InfoNCE
\cite{2018Oord} and MINE \cite{2018Belghazi}, are based on
the Donsker--Varadhan inequality for the KL divergence \cite{1983Donsker}.
This inequality states that

\begin{equation}
  D_\mathrm{KL}[P(x) \Vert Q(x)] \geq \mathbb{E}_{P(x)}[g(x)] - \ln
  \mathbb{E}_{Q(x)}[e^{g(x)}]
\end{equation}
for any bounded function $g(x)$ (i.e., there exists some $\xi <
\infty$ such that $|g(x)|<\xi$ for all $x$). The mutual information
$I(S;X)$ between two random variables $S$ and $X$ can be written as a
KL divergence
\begin{equation}
  I(S;X) = D_\mathrm{KL}[P(s,x) \Vert P(s)P(x)]
\end{equation}
between the joint distribution and the product of the marginal distributions.
Therefore, the Donsker--Varadhan inequality implies a lower bound for
the mutual information:
\begin{equation}\label{eq:SM-dv-ineq}
  I(S;X) \geq \mathbb{E}_{P(s,x)}[g(s,x)] - \ln
  \mathbb{E}_{P(s)P(x)}[e^{g(s,x)}]\,.
\end{equation}
Using this bound the mutual information can be estimated via
variational optimization of the scalar function $g(s, x)$ with the
goal of making the bound as tight as possible.

\subsection{Neural Variational Estimators: MINE and InfoNCE}

Neural variational MI estimators represent the function $g(s, x)$
via a neural network $g_\theta(s, x)$ and maximize the DV lower
bound. The neural network $g_\theta(s, x)$ is commonly called the
\emph{critic} with the optimal critic being
\begin{equation}\label{eq:SM-gstar}
  g^\star(s,x) = \ln\frac{P(s,x)}{P(s)P(x)}\,.
\end{equation}
Estimating the mutual information thus requires training the critic
such that $g_\theta(s,x)$ approximates the target $g^\star(s,x)$ as
closely as possible.
This is typically done via gradient ascent optimization of the
parameters $\theta$ to maximize the RHS of \cref{eq:SM-dv-ineq}.
The main difficulty lies in computing the normalization constant
\begin{equation}
  \label{eq:SM-z}
  Z = \mathbb{E}_{P(s)P(x)}[e^{g(s,x)}]
\end{equation}
which appears in \cref{eq:SM-dv-ineq}.

The normalization constant \cref{eq:SM-z} is an expectation over the
product distribution and cannot be computed in closed form;
MINE~\cite{2018Belghazi}
and InfoNCE~\cite{2018Oord} differ in how they approximate it.
For convenience, both MINE and InfoNCE introduce the notion of a
normalized critic $\hat r(s,x) = e^{g(s,x)}/Z$,
whose optimum, corresponding to the optimal critic $g^\star(s,x)$
of \cref{eq:SM-gstar}, is the density ratio $P(s,x)/[P(s)P(x)]$. Using this
normalized critic, the DV bound of \cref{eq:SM-dv-ineq} can be
rewritten in the simpler form
\begin{equation}
  \label{eq:SM-song-ermon}
  I(S;X) \geq \mathbb{E}_{P(s,x)}[\ln\hat{r}(s,x)]
\end{equation}
which holds for any function $\hat{r}(s,x)$ normalized such that
$\mathbb{E}_{P(s)P(x)}[\hat{r}(s,x)]=1$. For a derivation, see
Ref.~\cite{song2020understanding-e06}.

MINE \cite{2018Belghazi} estimates $Z$ via a direct Monte Carlo
average over $N$
independent samples $(s_i',x_i')$ drawn from the product
distribution $P(s)P(x)$. The resulting normalized critic of MINE is
\begin{equation}\label{eq:SM-mine}
  \hat r_\mathrm{MINE}(s,x) =
  \frac{e^{g(s,x)}}{\tfrac1N\sum_{i=1}^N e^{g(s_i',x_i')}}\,.
\end{equation}
While this Monte Carlo estimate of $Z$ is in principle unbiased,
the variance is potentially very large for high-dimensional $s$ and $x$.

InfoNCE \cite{2018Oord} instead normalizes per sample $s_i$, using
the other members of a batch of $N$ joint pairs as
negatives,
\begin{equation}\label{eq:SM-nce}
  \hat r_\mathrm{NCE}(s_i,x_i) =
  \frac{e^{g(s_i,x_i)}}{\tfrac1N\sum_{j=1}^N e^{g(s_i,x_j)}}\,.
\end{equation}
From this formula it is easy to see that unlike MINE the resulting
estimate $I_{\mathrm{NCE}} = \mathbb{E}_{P(s,x)}[\ln\hat
r_\mathrm{NCE}(s,x)]$ is
bounded from above by $\ln N$ \cite{2018Oord,2019Poole}.

In both cases the normalization constant $Z$ is computed via a
brute-force Monte Carlo average which is exact only as
$N\to\infty$. For finite $N$ and
high-dimensional data the integrand $e^{g(s,x)}$ is sharply peaked,
such that the Monte-Carlo computation of the normalization constant
becomes very difficult. The variance of the brute-force MC estimate of
$\mathbb{E}_{P(s,x)}[\ln\hat r_\mathrm{NCE}(s,x)]$ is therefore very
large. Crucially, the computation of the normalization constant
forms part of the training objective itself: the Monte Carlo variance
therefore affects not only the final estimate but the gradients
used to optimize the
critic, so that for insufficient $N$ training becomes unstable.
This is a well-understood
limitation of neural variational estimators
\cite{2019Poole,song2020understanding-e06}.

\subsection{ML-PWS: Sequential Monte Carlo Marginalization}

ML-PWS does not directly train a critic but instead trains a
generative model to estimate the conditional density $Q(x|s)$.
In a separate step, it computes a Monte-Carlo estimate of the marginal
\begin{equation}
  Q(x) = \mathbb{E}_{P(s)}[Q(x|s)]
\end{equation}
which corresponds to the critic
\begin{equation}
  \hat{r}_{\text{ML-PWS}}(s,x) = Q(x|s)/Q(x) \,.
\end{equation}
This ratio satisfies the constraint
$\mathbb{E}_{P(s)P(x)}[\hat{r}_{\text{ML-PWS}}(s,x)]=1$ required for
the validity of \cref{eq:SM-song-ermon}:
\begin{align}
  \mathbb{E}_{P(s)P(x)}\!\left[\frac{Q(x|s)}{Q(x)}\right]
  &= \mathbb{E}_{P(x)}\!\left[\frac{\mathbb{E}_{P(s)}[Q(x|s)]}{Q(x)}\right]
  = \mathbb{E}_{P(x)}\!\left[\frac{Q(x)}{Q(x)}\right]
  = 1\,.
\end{align}
Analogous to the other variational lower bounds, the ML-PWS lower
bound is given by
\begin{equation}
  I_{\text{ML-PWS}}(S;X) =
  \mathbb{E}_{P(s,x)}[\ln\hat{r}_{\text{ML-PWS}}(s,x)] \,.
\end{equation}

The qualitative difference between ML-PWS and the variational
estimators therefore lies in how the relevant normalization
constant is computed. As explained above, variational methods like MINE
compute $Z$ by brute-force
averaging the unnormalized score $e^{g(s'_i,x'_i)}$ over independent
draws from $P(s)\,P(x)$ and have no way to focus the
marginalization on inputs likely to have produced a
given output. ML-PWS, in contrast, exploits the assumption of the
availability of the input distribution $P(s)$ from which samples
can be drawn. This access to
$P(s)$ is precisely what makes a sequential particle filter
applicable to the marginalization (RR-PWS,
see~\cref{sec:SM-marginalisation}): the filter steers samples toward
the high-likelihood region of $P(s)\,Q(x|s)$ for each output
trajectory, instead of relying on independent draws from the prior $P(s)$.
The result is a much lower-variance estimate of $Q(x)$, and hence a
much smaller bias of the lower-bound estimator than the
brute-force sampling underlying the variational estimators.

% ============================================================
% SUPPLEMENTARY MATERIAL — Summary of Differences
% ============================================================

\section{Summary of Differences from Other Estimators}
\label{sec:SM-comparison-summary}

The preceding sections analyze in detail how ML-PWS relates to the
other machine-learning estimators of the mutual information: the neural
variational estimators MINE, InfoNCE, and SMILE (\cref{sec:SM-dv-bound}),
and the difference-of-entropies estimator DoE
(\cref{sec:SM-marginalisation}). For the reader's convenience we collect
here the principal differences.

All of these methods use machine learning to analyze high-dimensional
data, and all except DoE provide a lower bound on the mutual
information. What sets ML-PWS apart is that it learns only the
conditional distribution $Q(x|s)$ and then exploits the \emph{known}
input distribution $P(s)$ to marginalize it using efficient MC
trajectory-sampling techniques \cite{2023Reinhardt}, rather than relying on a
second learned model (as DoE does) or on a direct Monte Carlo average over a
finite batch (as the contrastive estimators do).

\paragraph{ML-PWS vs.\ DoE.}
Both ML-PWS and DoE train a generative model of the conditional
distribution $P(x_{1:n}|s_{1:n})$; they differ in how they obtain the
marginal $P(x_{1:n})$:
\begin{itemize}
  \item \emph{No second model.} DoE trains a \emph{separate} neural
    network to represent the marginal $P(x_{1:n})$, whereas ML-PWS
    obtains the marginal from the conditional by exact marginalization
    over $P(s_{1:n})$ (\cref{sec:SM-marginalisation}).
  \item \emph{Lower-bound guarantee.} Because the marginal is computed
    exactly rather than learned, ML-PWS
    yields a strict lower bound on the mutual information even for
    finite training data $N_\text{train}$ to develop the conditional model for
    $Q(x|s)$; the bias in the marginalisation of $Q(x|s)$ can be removed
    by extrapolating to the limit of infinite number of input trajectories $M$
    (\cref{sec:SM-marginalisation}); DoE offers no such guarantee and can
    over- or under-estimate the MI.
  \item \emph{No marginal-model error.} For nonlinear systems and
    systems with multiplicative noise the marginal is typically harder
    to learn than the conditional, so errors in DoE's marginal model
    bias its estimate. In the latent-factor benchmark
    (\cref{sec:SM-vmm-benchmark}) DoE's Gaussian marginal cannot capture
    the heavy-tailed true marginal and overestimates the MI, whereas
    ML-PWS remains accurate.
  \item \emph{Varying input statistics.} Since the marginal is computed
    analytically from the conditional for \emph{any} input
    distribution, ML-PWS handles non-stationary or otherwise modified
    inputs without retraining, and can be iterated to optimize the
    stimulus statistics and thereby estimate the channel capacity
    (\cref{sec:SM-capacity}); this is not possible with DoE.
\end{itemize}

\paragraph{ML-PWS vs.\ the variational estimators (MINE, InfoNCE,
SMILE).}

These estimators
\cite{2018Oord,2018Belghazi,song2020understanding-e06} train a neural
scoring function, or \emph{critic}, to distinguish samples from the
joint distribution $P(s,x)$ from samples drawn from the product distribution
$P(s)P(x)$, in order to estimate the density ratio
$r(s,x)=P(x|s)/P(x)$, see \cref{sec:SM-dv-bound}. ML-PWS instead uses the
$N_\mathrm{train}$ measured input--output pairs to directly learn a
generative model $Q(x|s)$ of the conditional output
distribution, and marginalizes it against the known input
distribution to obtain the required density ratio:
\begin{itemize}
  \item \emph{No $\ln N$ saturation.} In InfoNCE the critic must
    identify, among $N$ candidates consisting of one positive and
    $N-1$ negative samples, the output trajectory matching a given
    input trajectory. As the trajectories become longer, the correct
    input--output pairing generally becomes increasingly
    distinguishable from the mismatched pairings, and the probability
    of selecting the correct output can approach unity. At that point
    the classification task is solved and the objective saturates:
    each of the $N$ possible pair identities is perfectly resolved,
    and, because these identities are sampled with equal prior weight,
    their entropy is $\ln N$. The InfoNCE estimate therefore cannot
    exceed $\ln N$ nats. In contrast, the ML-PWS estimate is
    \emph{not} obtained
    from an $N$-way classification problem and
    has no such intrinsic ceiling: the finite data set limits the
    accuracy of the learned model $Q(x|s)$ and thus the tightness
    of the bound,
    but it does not impose an
    upper bound on the resulting estimate.
  \item \emph{Bias--variance trade-off at large MI.} MINE and SMILE do
    not possess this same explicit $\ln N$ ceiling, but they face the
    related difficulty that estimating large density ratios
    $r(s,x)=P(s,x)/P(s)P(x)$ via brute-force
    averages over $P(s)P(x)$ leads to large variance; SMILE reduces
    this variance by
    clipping the density ratio at $\tau$, at the cost of a bias that
    caps the estimate at $2\tau$. The critic-based estimators
    therefore suffer from an unfavourable bias--variance and
    sample-complexity trade-off precisely in the large-information
    regime that is relevant for computing the information rate.
  \item \emph{Marginalization error controlled independently of the
    data.} ML-PWS exploits the fact that $P(s)$ is known from the
    experimental protocol and exactly computes the corresponding marginal
    $Q(x) = \mathbb{E}_{P(s)}[Q(x|s)]$ via Path Weight Sampling.
    PWS leverages
    trajectory-sampling methods developed for free-energy calculations
    in statistical physics and for polymer sampling: rather than
    propagating all input trajectories independently over the full
    duration, it generates an ensemble of $M$ trajectories
    collectively, segment by segment, and uses reweighting and
    resampling to concentrate the computational effort on the input
    histories that contribute most strongly to the marginal
    (\cref{sec:SM-marginalisation}). This is a vital ingredient: it makes
    accurate marginalization feasible at moderate $M$.
    Since $M$ is a computational parameter independent of the dataset size
    $N_\mathrm{train}$, it can be made arbitrarily large;
    by extrapolating the estimate to
    $M\to\infty$, the marginalization of the learned conditional model
    can be performed to arbitrarily high accuracy (\cref{sec:SM-bias}).
    The contrastive estimators have no analogue of this: their
    normalization is a brute-force average over $P(s)P(x)$, which has
    high variance in high dimensions and whose accuracy is tied to the
    batch size $N \leq N_\mathrm{train}$.
\end{itemize}

% ============================================================
% SUPPLEMENTARY MATERIAL — Efficient Marginalisation
% ============================================================

\section{Efficient Marginalization}
\label{sec:SM-marginalisation}

The central computational challenge in ML-PWS is evaluating the
marginal path probability
\begin{equation}
  Q(x_{1:n}) = \mathbb{E}_{\prob(s_{1:n})}\bigl[Q(x_{1:n}|s_{1:n})\bigr]
  \label{eqn:SM-marginal_def}
\end{equation}
for each observed output trajectory $x_{1:n}$. This marginalization
is what distinguishes ML-PWS from DoE: it is an exact expectation
over the known input distribution $\prob(s_{1:n})$, rather than the
output of a second trained model for the marginal probability.

The na\"{i}ve Monte Carlo estimate of \cref{eqn:SM-marginal_def},
\begin{equation}
  \hat{Q}(x_{1:n}) = \frac{1}{M}\sum^M_{j=1}Q(x_{1:n}|s^j_{1:n})
  \label{eqn:SM-direct_mc}
\end{equation}
with input trajectory samples $s^j_{1:n}$ drawn from the input
distribution $P(s_{1:n})$,
becomes exponentially inefficient as the trajectory length $n$
increases. For long trajectories,
almost all weight is concentrated on a vanishingly small fraction of
the $M$ samples, and the effective sample size $M_\text{eff}$ falls
far below $M$.
We discuss two complementary strategies to overcome this problem.

\begin{figure}
  \centering
  \includegraphics[width=\linewidth]{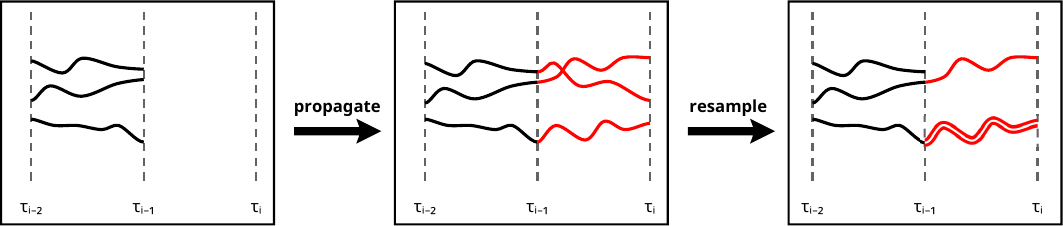}
  \caption{The sequential Monte Carlo scheme used in RR-PWS (Figure
    taken from Ref.~ \cite{2023Reinhardt}) to generate an
    ensemble of input trajectories $s_{0:i}$ for a given output
    trajectory $x_{0:i}$.
    We start with a set of trajectories $s^j_{0:i-1}$ with time span
    $[\tau_0,\tau_{i-1}]$ (left panel). In the next step we propagate
    these trajectories forward in time to $\tau_i$, according to
    $\prob(s_{i}\mid s^k_{0:i-1})$ (central panel). Then we resample
    the trajectories according to the Boltzmann weights of their most
    recent segments, effectively pruning or enriching individual
    segments. An example outcome of the resampling step is shown in the
    right panel where the bottom trajectory was duplicated and one of
    the top trajectories was eliminated. These steps are repeated for
    each segment, until a set of input trajectories of the desired
    length is generated. The intermediate resampling steps bias the
    trajectory distribution from $\prob(s_{0:i})$ towards
  $\prob(s_{0:i}|x_{0:i})$.}
  \label{fig:SM-FigureSMC}
\end{figure}

\textbf{Strategy 1: Sequential Monte Carlo (RR-PWS).}
For each output trajectory $\mathbf{x}$, RR-PWS \cite{2023Reinhardt}
performs the marginalization by generating the $M$ input trajectories
collectively, segment by segment, rather than by propagating them
independently over the full duration.
While the full algorithm is described in detail in
Ref.~\cite{2023Reinhardt}, we provide a brief overview of the main steps here.

The total duration of the trajectories is divided into $K$
segments of length $\ell = n/K$, and
the marginal is estimated by propagating an
ensemble of $M$ input trajectories for each output trajectory.
In all simulations reported in
this work we use the finest possible segmentation, one segment per
discrete time step, so that $\ell=1$ and $K=n$.
Along with the partial trajectories, the algorithm keeps track of a
set of weights $\bar{W}^{\,j}_{k}$ per trajectory segment to
perform the marginalization.
Specifically, let $\bar{W}^{\,j}_{k-1}$
denote the normalized weight carried into segment $k$ by trajectory $j$,
with the uniform initialization $\bar{W}^{\,j}_0 = 1/M$.
At step $k$,
each trajectory is extended by drawing a new segment
$s^j_{(k-1)\ell+1:k\ell}$ from the input prior
$\prob(s_{(k-1)\ell+1:k\ell}\mid s^j_{1:(k-1)\ell})$ and we compute the
incremental weight of this segment as
\begin{equation}
  w^j_k = Q\!\left(x_{(k-1)\ell+1:k\ell}
  \,\big|\, s^j_{1:k\ell},\, x_{1:(k-1)\ell}\right).
  \label{eq:SM-incremental_weight}
\end{equation}
The corresponding marginal factor of the segment,
$Q(x_{(k-1)\ell+1:k\ell}\mid x_{1:(k-1)\ell})$, is estimated by the
weighted ensemble mean of the incremental weights,
\begin{equation}
  \hat{Q}\!\left(x_{(k-1)\ell+1:k\ell}\mid x_{1:(k-1)\ell}\right)
  = \sum^M_{j=1}\bar{W}^{\,j}_{k-1}\,w^j_k \,.
  \label{eq:SM-segment_factor}
\end{equation}

Now, before proceeding to the next segment, we potentially
perform a \emph{resampling} step, based on an adaptive resampling criterion.
If the resampling criterion is triggered, a new ensemble of $M$
trajectories is drawn with replacement using probabilities
\begin{equation}
  \pi^{\,j}_k = \frac{\bar{W}^{\,j}_{k-1}\,w^j_k}
  {\sum_{i=1}^M \bar{W}^{\,i}_{k-1}\,w^i_k}
  \,,
  \label{eq:SM-weight_update}
\end{equation}
i.e., we sample $M$ times from the pool
$\{s^1_{1:k\ell},\ldots,s^M_{1:k\ell}\}$ such that each time trajectory
$s^j_{1:k\ell}$ is selected with probability $\pi^{\,j}_k$.
This effectively prunes the low-weight
trajectories that contribute negligibly to the marginal in favor of copies of
the high-weight ones.
After resampling is performed, the weights $\bar{W}^{\,j}_k$ of all
trajectories are reset to the uniform value $1/M$.
In contrast, when the resampling criterion is not triggered, the
ensemble of trajectories is left unchanged and we set the
normalized weights to $\bar{W}^{\,j}_k = \pi^{\,j}_k$.
In our implementation, the resampling criterion is chosen such that
resampling at segment $k$ is
triggered whenever the effective sample size
$M_\text{eff} = \left(\sum_j \pi^j_k\right)^2 /
\sum_j(\pi^j_k)^2$ is below $M/2$, following the recommendation of
Ref.~\cite{2017Martino}. The SMC procedure is also depicted
graphically in \cref{fig:SM-FigureSMC}.

Since the marginal is the product of the segment factors given in
\cref{eq:SM-segment_factor}, the full
marginal is given by
\begin{equation}
  \ln\hat{Q}(x_{1:n})
  = \sum^K_{k=1}\ln\!\left(\sum^M_{j=1}\bar{W}^{\,j}_{k-1}\,w^j_k\right).
  \label{eqn:SM-rr_pws}
\end{equation}
While the SMC estimator $\hat{Q}(x_{1:n})$ itself is unbiased,
its logarithm \emph{is} biased due to Jensen's inequality.
The empirical scaling of
the resulting marginalisation bias with trajectory length $n$ and
number of propagated trajectories $M$ is documented in \cref{sec:SM-bias}.

\textbf{Strategy 2: Importance sampling via a trained posterior
approximation.}
For a fixed output trajectory $x_{1:n}$, the importance-sampling
proposal that drives the Monte Carlo variance of $\hat{Q}(x_{1:n})$
to zero is the model posterior
$Q(s_{1:n}|x_{1:n})\propto\prob(s_{1:n})\,Q(x_{1:n}|s_{1:n})$.
Since this posterior is not
available in closed form, we train an autoregressive inference
network $\mathcal{Q}(s_{1:n}|x_{1:n})$ to approximate it by
minimizing the Kullback--Leibler divergence
$D_\mathrm{KL}[\mathcal{Q}(s|x)\Vert Q(s|x)]$, equivalently
maximizing the Evidence Lower Bound (ELBO) \cite{2013Kingma}
\begin{equation}
  \mathrm{ELBO}(x_{1:n}) =
  \mathbb{E}_{\mathcal{Q}(s_{1:n}|x_{1:n})}
  \!\left[
    \ln\frac{\prob(s_{1:n})Q(x_{1:n}|s_{1:n})}{\mathcal{Q}(s_{1:n}|x_{1:n})}
  \right]\!.
  \label{eqn:SM-elbo}
\end{equation}
The importance-sampled estimate of the marginal then becomes
\begin{equation}
  \hat{Q}(x_{1:n}) = \frac{1}{M}\sum^M_{j=1}
  \frac{\prob(s^j_{1:n})\,Q(x_{1:n}|s^j_{1:n})}{\mathcal{Q}(s^j_{1:n}|x_{1:n})}.
  \label{eqn:SM-is_mc}
\end{equation}
where the $s^j_{1:n}$ are drawn from the inference network
$\mathcal{Q}(s_{1:n}|x_{1:n})$.
This estimate is unbiased for any choice of $\mathcal{Q}$ that
shares the support of $\prob(s_{1:n})\,Q(x_{1:n}|s_{1:n})$, and its
variance decreases as $\mathcal{Q}$ approaches the model posterior.
In the limit $\mathcal{Q}(s|x)\to Q(s|x)$ all importance weights become
equal and the variance vanishes. $\mathcal{Q}$ can be implemented as an
autoregressive normalising flow \cite{2017Papamakarios} and trained
jointly with the generative model $Q(x|s)$.

\textbf{Combining both strategies.}
The two strategies are complementary and can be combined within the
RR-PWS framework. The modification is straightforward: instead of
drawing each segment $k$ from the prior
$\prob(s_{(k-1)\ell+1:k\ell}|s^j_{1:(k-1)\ell})$, we draw it from
the posterior approximation
$\mathcal{Q}(s_{(k-1)\ell+1:k\ell}|s^j_{1:(k-1)\ell},x_{1:n})$.
The incremental importance weight is then corrected accordingly,
\begin{equation}
  w^j_k = \frac{\prob(s^j_{(k-1)\ell+1:k\ell}|s^j_{1:(k-1)\ell})\,
  Q(x_{(k-1)\ell+1:k\ell}|s^j_{1:k\ell},\,x_{1:(k-1)\ell})}
  {\mathcal{Q}(s^j_{(k-1)\ell+1:k\ell}|s^j_{1:(k-1)\ell},x_{1:n})},
  \label{eq:SM-combined_weight}
\end{equation}
and $\hat{Q}(x_{1:n})$ is again given by \cref{eqn:SM-rr_pws}.
This combined approach remains bias-free for computing
$\hat{Q}(x_{1:n})$. Segments drawn from $\mathcal{Q}$ are concentrated
in the high-posterior region, reducing weight collapse at each
resampling step and therefore requiring fewer trajectories $M$ to achieve
a given estimation accuracy. The standard RR-PWS and the pure ELBO
approaches of Strategies 1 and 2 are recovered as special cases when
$\mathcal{Q}(s|x)=\prob(s)$ and $K=1$, respectively.
All results reported in this work use the standard
RR-PWS scheme (Strategy 1) exclusively: the importance-sampling
extension (Strategy 2) and the combined scheme are described here for
completeness but are not employed in any of our experiments.

\section{Empirical Evaluation of Marginalization Bias}\label{sec:SM-bias}

The lower bound \eqref{eq:SM-gap} holds under the condition that
$Q(x_{1:n})$ is the exact marginal of $Q(x_{1:n}|s_{1:n})$ under
$\prob(s_{1:n})$. Yet, in practice, the marginal is numerically evaluated by a
Monte Carlo estimator $\hat{Q}_M(x_{1:n})$ that averages importance
weights over $M$ trajectories drawn from a proposal distribution.
While $\hat{Q}_M(x_{1:n})$ is an unbiased estimator of $Q(x_{1:n})$, the
estimator $-\ln\hat{Q}_M$ is biased due to Jensen's inequality. This
section quantifies that bias and shows how
to separate it from the asymptotic estimator by a $1/M$
extrapolation.
% , and reports the empirical finding that the bias is
% controllable for the SMC-based marginalization of PWS but not for
% the brute-force Monte Carlo marginalization of the same generative
% model.

\textit{Leading-order bias of the finite-$M$ estimator.}---We write the Monte
Carlo estimator for the marginal path likelihood as
\begin{equation}
  \hat{Q}_M(x_{1:n}) = \frac{1}{M}\sum_{i=1}^M w_i(x_{1:n})\,,
  \label{eqn:SM-QM}
\end{equation}
where $w_i(x_{1:n})$ are the importance weights of the full
trajectory $x_{1:n}$.
For brute-force marginalization, $w_i(x_{1:n}) =
Q(x_{1:n}|s^i_{1:n})$ for iid samples $s^i_{1:n}$ drawn from $\prob(s_{1:n})$.
For RR-PWS, the full trajectory weight is the product of the $K$
incremental segment weights defined in \cref{eq:SM-incremental_weight},
\begin{equation}
  w_i(x_{1:n}) = \prod_{k=1}^K w^i_k\,.
  \label{eqn:SM-full_weight}
\end{equation}

The entropy estimator is built
from $-\ln \hat{Q}_M$, and a second-order Taylor expansion around $Q$
gives
\begin{equation}
  \mathbb{E}\!\left[-\ln \hat{Q}_M(x_{1:n})\right]
  = -\ln Q(x_{1:n})
  + \frac{1}{2M}\,\frac{\mathrm{Var}[w_1(x_{1:n})]}{Q(x_{1:n})^2}
  + O(M^{-2})\,,
  \label{eqn:SM-jensen_expansion}
\end{equation}
where $\mathrm{Var}[w_1]$ is the variance of a single importance weight
(the same for all $i$ by independence).
Averaging over the data
distribution yields the bias in the marginal entropy estimate
\begin{equation}
  \hat{H}_M(X_{1:n}) - \hat{H}_\infty(X_{1:n})
  = \frac{c(n)}{M} + O(M^{-2})\,,\quad
  c(n) = \frac{1}{2}\!\left\langle
  \frac{\mathrm{Var}[w_1(x_{1:n})]}{Q(x_{1:n})^2}\right\rangle_{\prob(x_{1:n})}\geq
  0\,.
  \label{eqn:SM-H_bias}
\end{equation}

The estimator for the conditional entropy $\hat{H}(X_{1:n}|S_{1:n})$
evaluates $Q(x|s)$ directly from the learned network and carries no
marginalization step, so it is not biased. The mutual
information estimate therefore directly inherits the marginal-entropy bias:
\begin{equation}
  \hat{I}_M = \hat{I}_\infty + \frac{c(n)}{M} + O(M^{-2})\,.
  \label{eqn:SM-mi_bias}
\end{equation}
At finite $M$, the estimate is larger than $\hat{I}_\infty$; the
lower bound is therefore asymptotic in $M$.
For brute-force marginalization $w =
Q(x_{1:n}|s_{1:n})$ under $s_{1:n}\sim\prob(s_{1:n})$, and this
variance grows rapidly with trajectory length: nearly all $s_{1:n}$
drawn from the prior produce negligible likelihood for a typical
$x_{1:n}$, and the weight distribution becomes heavy-tailed. A
proposal $\tilde{Q}(s_{1:n}|x_{1:n})$ that tracks the
posterior $\prob(s_{1:n}|x_{1:n})$ instead yields a nearly constant
weight, with $\mathrm{Var}[w_1(x_{1:n})]=0$ in the limit
$\tilde{Q}=\prob(s_{1:n}|x_{1:n})$; the
SMC scheme of RR-PWS approximates this ideal by updating the weights
sequentially in time, and the resulting scaling of
$\mathrm{Var}[w_1(x_{1:n})]$ with $n$ is
quantified empirically below.

\textit{Diagnostic via $1/M$ extrapolation.}---\cref{eqn:SM-mi_bias}
suggests a direct diagnostic: compute $\hat{I}_M$ for several values
of $M$, plot the estimate against $1/M$, and read the intercept at
$1/M\to 0$ as the bias-corrected estimate of $\hat{I}_\infty$. The
slope of the linear fit is a direct measurement of $c(n)$ and the
goodness of the fit serves as an indication of whether the sampling
regime is asymptotic.

\begin{figure}
  \centering
  \includegraphics[width=\columnwidth]{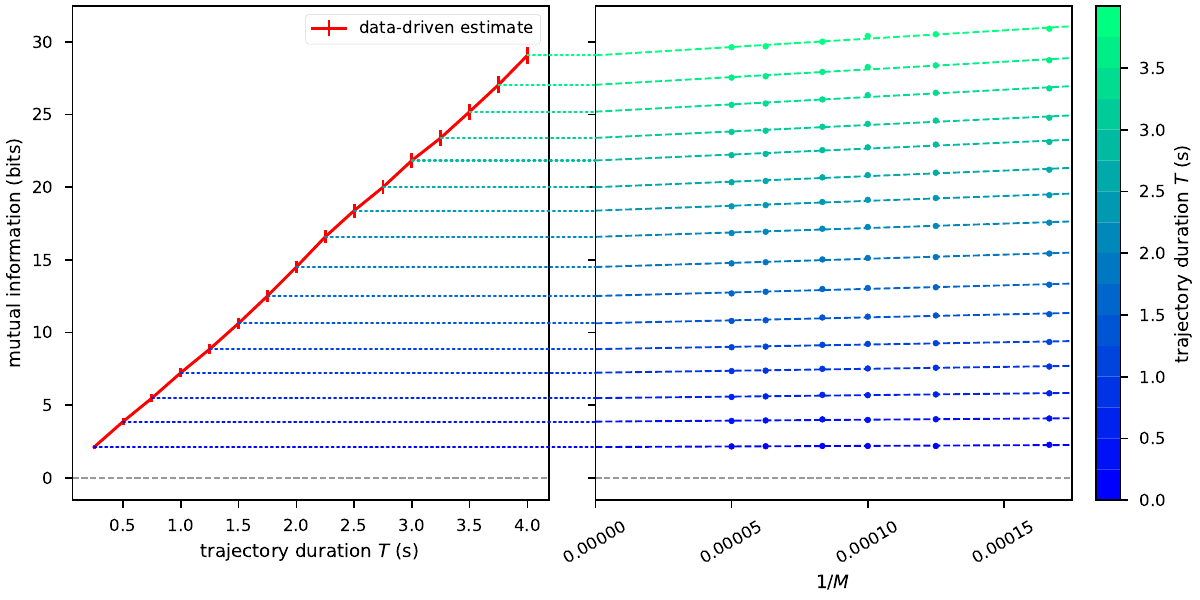}
  \caption{
    Empirical analysis of the marginalization bias for the
    neural population model. (a)
    Mutual information estimate $\hat{I}_M(T)$ as a function of
    trajectory duration $T$. The red curve is the bias-corrected
    estimate obtained by linearly extrapolating
    $1/M\to 0$, with error bars showing the across-trial standard
    error ($N=830$ trajectories).
    (b) The same data plotted against
    $1/M$ for each $T$ (colour-coded). Dashed lines are
    linear fits of $\hat{I}_M$ in $1/M$; the intercept at $1/M=0$ is
    the bias-corrected estimate shown in panel~(a).
  }
  \label{fig:SM-marginal_bias}
\end{figure}

\textit{Example for the neural population model.}---We apply this
diagnostic to the trained model of the $50$-neuron population. The
conditional likelihood $Q(x_{1:n}\mid s_{1:n})$ is the Poisson spiking
likelihood of the network, evaluated exactly on the paired data, while
the stimulus $\prob(s_{1:n})$ is given by a stochastic
harmonic-oscillator process; both are time-discretized with $\Delta t =
10\,\mathrm{ms}$ (for more details on the neuronal system, see
\cref{sec:SM-ganglion}). For a given number of SMC trajectories $M$, the marginal
$\hat{Q}_M(x_{1:n})$ of \cref{eqn:SM-QM} is computed by the RR-PWS scheme: $M$
stimulus trajectory segments are drawn from the prior and advanced in
time with a
bootstrap particle filter that accumulates the segment weights of
\cref{eqn:SM-full_weight} and resamples adaptively whenever the effective
sample size drops below $M/2$ (see \cref{fig:SM-FigureSMC}). For each of the
$N$ recorded spike trajectories $x_{1:n}$ we subtract the SMC
marginal from the cumulative conditional log-likelihood, yielding the
running estimate $\hat{I}_M(T)$ as a function of trajectory duration
$T=n\Delta t$. This is repeated for various values of $M$.

Before running the estimator on an output trajectory, we \emph{warm up} the
network with the \SI{500}{\milli\second} of true stimulus that precede
that trajectory in the recording, instead of initializing the stimulus history
with zeros. This is helpful because the spikes fired at the
beginning of a trajectory are driven in part by the stimulus that came
before it, over the \SI{290}{\milli\second} receptive field of the
artificial neural network. Because the network is conditioned on
this past, the
quantity we estimate is thus the mutual information \emph{conditioned} on
the stimulus history. At short durations this differs from the
unconditioned mutual information, but the difference is a boundary
contribution, only relevant for small $T$: the two grow with the same slope,
so the information rate can be correctly estimated.

We use the same warm-up stimulus for the conditional likelihood
$Q(x_{1:n}|s_{1:n})$ and for the $M$ stimulus trajectories sampled in
the SMC marginalization, which are all continued from that common past,
so that the estimate remains a valid lower bound.
The results are shown in \cref{fig:SM-marginal_bias}.
We find that the MI estimates are linear in $1/M$, and hence
the intercept of the trajectory means can be viewed
as a reliable estimate of the mutual information lower bound.

\section{Example I: A Simple Nonlinear Model with AR(3) Input}
\label{sec:SM-example1}

Example~I is the simplest of the three benchmarks and is presented
here rather than in the main text. It has the same qualitative
structure as Example~II below (\cref{eq:SM-nonlinear_model}): a linear
autoregressive input combined with a saturating, stochastic, scalar
output. The input is an $\mathrm{AR}(p)$ process,
\begin{equation}
  S_t = \sum^p_{j=1} \phi_j S_{t-j} + \xi_t,
  \label{eq:SM-AR1_other}
\end{equation}
with $\xi_t\sim\mathcal{N}(0,1)$ iid and $\phi_j\in[0,1)$. The
output is
\begin{equation}
  X_t = \sigma(\gamma S_t) + \rho X_{t-1} + \vartheta \eta_t,
  \label{eq:SM-nonlinear_model_other}
\end{equation}
with $\sigma(x)=1/(1+e^{-x})$ the logistic function and
$\eta_t\sim\mathcal{N}(0,1)$ iid. We use $p=3$ with parameter values
listed in \cref{tab:SM-model_parameters}; for each value of the
nonlinearity gain $\gamma$ we generate a training set of $N=1000$
pairs $(s_{1:50}, x_{1:50})$. The generative network is a Gaussian
LSTM trained by maximum likelihood; full hyperparameters are
analogous to those of \cref{sec:SM-nonlinear-training}, and PWS
marginalization uses the RR-PWS scheme of
\cref{sec:SM-marginalisation}.

\begin{table}
  \centering
  \caption{Model Parameters for
    \cref{eq:SM-AR1_other,eq:SM-nonlinear_model_other}. We use an
  $\mathrm{AR}(3)$ process.}
  \label{tab:SM-model_parameters}
  \begin{tabular}{ccccccc}
    % \toprule
    & \multicolumn{3}{c}{\textbf{Input}} &
    \multicolumn{3}{c}{\textbf{Output}} \\
    \cmidrule(lr){2-4} \cmidrule(lr){5-7}
    \textbf{Parameter} & $\phi_1$ & $\phi_2$ & $\phi_3$ & $\gamma$ &
    $\rho$ & $\vartheta$ \\
    \midrule
    \textbf{Value} & 0.5 & -0.3 & 0.2 & 1.0 & 0.2 & 0.2 \\
    % \bottomrule
  \end{tabular}
\end{table}

\begin{figure*}
  \centering
  \includegraphics[width=\textwidth]{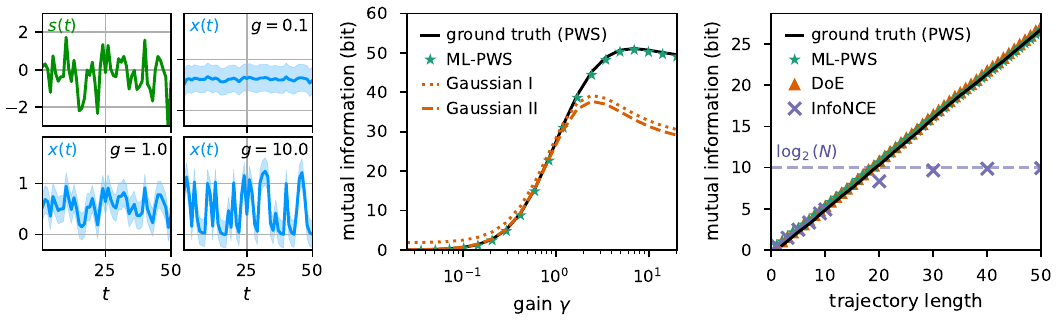}
  \caption{Test of ML-PWS against ground truth PWS result for the
    non-linear model of \cref{eq:SM-nonlinear_model_other} with an
    AR(3) input [\cref{eq:SM-AR1_other}]. (a)
    Example time series from the training set.
    Upper left panel: one stochastic realization of the input.
    Other panels: mean of the output distribution as well as 10/90-th
    percentiles, for different values of the gain $\gamma$.
    (b) Mutual information estimates as a function of gain $\gamma$.
    For low gain ($\gamma \lesssim 1$), both Gaussian approximations
    align closely with the ground truth and ML-PWS. At high gain,
    however, the Gaussian approximation fails to capture nonlinear effects.
    In contrast, ML-PWS correctly estimates the mutual information for
    the full range of $\gamma$. (c) Comparison of ML-PWS against
    InfoNCE and DoE for the path mutual information as a function of
  trajectory length for the same non-linear model. We set $\gamma=1.0$.}
  \label{fig:SM-ARtest}
\end{figure*}

The results, summarised in \cref{fig:SM-ARtest}, are qualitatively the
same as for the systems presented in the main text. Panel~(b) shows that ML-PWS
tracks the PWS ground truth across the full range of $\gamma$, while
the Gaussian approximation deviates once $\gamma\gtrsim 1$ and the
logistic saturation becomes important; we report the Gaussian
estimate both at the training-set size $N=1000$ (Gaussian~I) and at
$N=10^5$ (Gaussian~II) to separate finite-sample bias from the
linearity bias. Panel~(c) compares ML-PWS to InfoNCE and DoE as a
function of trajectory length: ML-PWS and DoE remain accurate at all
lengths, while InfoNCE saturates at the $\ln N$ ceiling for long
trajectories \cite{2018Oord,2019Poole}.

\section{Example II: Nonlinear Model}
\label{sec:SM-example2}

To generate training data for the neural network, we combine a
linear stochastic input with a nonlinear output model. The input
$S_t$ has $\mathrm{ARMA}(2,1)$ statistics, which are described by the
canonical recurrence
\begin{equation}
  S_t = a_1S_{t-1} + a_2S_{t-2} + b_0\xi_t + b_1\xi_{t-1},
  \label{eq:SM-AR1}
\end{equation}
where $\xi_t \sim \mathcal{N}(0,1)$ are iid. Rather than specifying
the four ARMA coefficients directly, we parameterize the input $S_t$ as a
sum of two independent Ornstein--Uhlenbeck processes,
\begin{equation}
  \mathrm{d}u_j(t) = -\frac{u_j(t)}{\tau_j}\,\mathrm{d}t + \sigma_j\,
  \mathrm{d}W_j(t), \qquad j=1,2,
  \label{eq:SM-ou_components}
\end{equation}
with correlation times $\tau_1,\tau_2$ and noise intensities
$\sigma_1,\sigma_2$, and set $S(t) = u_1(t) + u_2(t)$. The
stationary variance of each component is $\sigma_j^2\tau_j/2$.
Sampling at step $\Delta t$ and using the matched-pole
(matrix-exponential) discretization, each component reduces to an
AR(1) process
\begin{equation}
  u_{j,t} = \phi_j\,u_{j,t-1} + \theta_j\,\epsilon_{j,t}, \qquad
  \epsilon_{j,t}\sim\mathcal{N}(0,1)\text{ iid},
\end{equation}
with
\begin{equation}
  \phi_j = e^{-\Delta t/\tau_j}, \qquad
  \theta_j^2 = \frac{\sigma_j^2\,\tau_j}{2}\,(1-\phi_j^2).
\end{equation}
As shown in \cref{sec:SM-ar1sum} below, the sum $S_t = u_{1,t} + u_{2,t}$
is not an AR($p$) process for any
$p > 0$, but a non-Markovian ARMA(2,1) process,
with coefficients $(a_1,a_2,b_0,b_1)$
determined by $(\phi_j,\theta_j)$. The values of $\tau_j$,
$\sigma_j$, and $\Delta t$ used in the experiments are listed in
\cref{tab:SM-nonlinear_dataset}.

\begin{figure}
  \centering
  \includegraphics[width=0.9\linewidth]{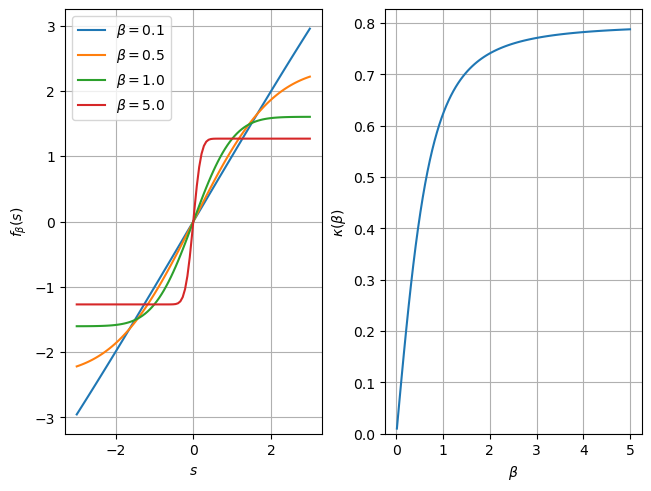}
  \caption{Nonlinearity $f_\beta$ (\cref{eq:SM-f_beta}) used in the
    first case study.
    \textit{Left:} The function $f_\beta(x)$ for several values of
    the nonlinearity
    strength $\beta$; small $\beta$ approaches the linear
    (identity) regime, while large
    $\beta$ approaches a step function.
    \textit{Right:} The normalization factor $\kappa(\beta)$, equal to the
    Gaussian-input gain
    $\mathbb{E}[S\,f_\beta(S)]/\mathbb{E}[S^2]$, as a function of
    $\beta$; dividing by $\kappa(\beta)$ ensures that
    $\mathrm{std}[\alpha f_\beta(S_t)] = \alpha$
  regardless of the value of $\beta$.}
  \label{fig:SM-f_Beta}
\end{figure}

The output $X_t$ is governed by the equation
\begin{equation}
  X_t = \alpha f_\beta(S_t) + \rho X_{t-1} + \vartheta \eta_t
  \label{eq:SM-nonlinear_model}
\end{equation}
where $\eta_t$ are iid Gaussian random numbers, and $\alpha$, $\rho$ and
$\vartheta$ are positive real parameters; throughout we fix the
output-noise amplitude to $\vartheta=1$. The nonlinearity
\begin{equation}
  f_\beta(x) = \frac{\erf\left(\frac{\sqrt{\pi}}{2} \beta
  x\right)}{\kappa(\beta)}
  \label{eq:SM-f_beta}
\end{equation}
where $\erf(x)=2/\sqrt{\pi} \int^x_0 dt\,e^{-t^2}$ is the error function and
\begin{equation}
  \kappa(\beta) = \frac{\beta}{\sqrt{1 + \frac{\pi}{2}\beta^2}}
\end{equation}
is the Gaussian-input gain
$\mathbb{E}[S\,f_\beta(S)]/\mathbb{E}[S^2]$ for the unit-variance
input $S_t$, included so that the standard deviation of the output signal,
$\mathrm{std}[\alpha\,f_\beta(S_t)]$, equals $\alpha$ regardless of the
nonlinearity strength $\beta$.
Indeed, $\beta$ effectively controls the strength of the nonlinearity, with
$\beta\to 0$ approaching a linear response and large $\beta$ producing
hard saturation.

The dataset parameters used to generate Fig.~1(b) and~(c) of the
main text are listed in \cref{tab:SM-nonlinear_dataset}. In each panel one
parameter is varied across $21$ discrete values while all
others are held fixed at the values listed in \cref{tab:SM-nonlinear_dataset}.

It is useful to relate the sequence length to the correlation time of
the input. With the time step $\Delta t = 0.1$, the discretized input
has correlation lengths $\tau_1/\Delta t = 5$ and $\tau_2/\Delta t =
25$ time steps for the fast and slow OU modes, respectively. The
sequence length $T=30$ therefore spans about six correlation times of
the fast mode but only $\approx 1.2$ correlation times of the slow
mode, so each trajectory contains roughly one independent fluctuation
of the slow input component. The training set consequently contains on the
order of $N_\mathrm{train}\,T/(\tau_2/\Delta t)\approx 1.2\times10^3$
effectively independent samples of the slow input mode.

\begin{table}
  \centering
  \caption{Dataset parameters for the nonlinear model
    [\cref{eq:SM-AR1,eq:SM-nonlinear_model}] used in Fig.~1(b) and~(c) of the
    main text. In each panel one parameter is swept while the others are
  held fixed; the swept range is given in braces.}
  \label{tab:SM-nonlinear_dataset}
  \begin{tabular}{llll}
    \textbf{Parameter} & \textbf{Symbol} & \textbf{Panel (b)} &
    \textbf{Panel (c)} \\
    \midrule
    \multicolumn{4}{l}{\textit{Output [\cref{eq:SM-nonlinear_model}]}} \\
    Output gain & $\alpha$ & $\{0.1,\, 0.5,\, 1.0,\, 1.5,\, \ldots,\,
    10.0\}$ & $10.0$ \\
    Nonlinearity slope & $\beta$ & $5.0$ & $\{0.1,\, 1.0,\, 2.0,\,
    \ldots,\, 20.0\}$ \\
    Time step & $\Delta t$ & $0.1$ & $0.1$ \\
    AR decay & $\rho = e^{-\Delta t}$ & $\approx 0.905$ & $\approx
    0.905$ \\
    Noise amplitude & $\vartheta$ & $1.0$ & $1.0$ \\
    \midrule
    \multicolumn{4}{l}{\textit{Input (sum of two OU processes,
    \cref{eq:SM-ou_components})}} \\
    OU correlation times & $\tau_1,\,\tau_2$ & $0.5,\, 2.5$ & $0.5,\, 2.5$ \\
    OU noise intensities & $\sigma_1,\,\sigma_2$ & $0.44,\, 0.20$ &
    $0.44,\, 0.20$ \\
    \midrule
    \multicolumn{4}{l}{\textit{Dataset}} \\
    Sequence length & $T$ & $30$ & $30$ \\
    Training trials & $N_\mathrm{train}$ & $1000$ & $1000$ \\
    Validation trials & $N_\mathrm{val}$ & $1000$ & $1000$ \\
  \end{tabular}
\end{table}

\subsection{Input Statistics: The Sum of Two AR(1) Processes}
\label{sec:SM-ar1sum}

We chose the input of \cref{eq:SM-AR1} because its correlations span two
distinct time scales, which makes the marginalization over the input
history non-trivial. The simplest way to obtain such a process is the
superposition of two AR(1) processes with different relaxation times,
\cref{eq:SM-ou_components}. Here we show that this superposition is not
an AR($p$) process for any $p>0$, but a non-Markovian ARMA(2,1)
process, i.e.\ that it obeys the recurrence \cref{eq:SM-AR1}, and we
derive the coefficients $(a_1,a_2,b_0,b_1)$ entering that recurrence in
terms of the AR(1) parameters $(\phi_j,\theta_j)$.

Let $u_1[n]$ and $u_2[n]$ be two independent AR(1) processes defined as
\begin{align}
  u_1[n] &= \phi_1 u_1[n-1] + \theta_1\epsilon_1[n] \\
  u_2[n] &= \phi_2 u_2[n-1] + \theta_2\epsilon_2[n] \,.
\end{align}
with independent white noise processes $\epsilon_1[n]$ and
$\epsilon_2[n]$ and real parameters
$\phi_1,\theta_1,\phi_2,\theta_2$. A generic ARMA($p$, $q$) process
$v[n]$ is defined via
\begin{equation}
  v[n] = \sum^p_{i=1} a_i v[n-i] + \sum^q_{i=0} b_i \epsilon[n-i]
\end{equation}
with real parameters $a_1,\ldots, a_p,b_0,\ldots,b_q$ and white
noise $\epsilon[n]$; \cref{eq:SM-AR1} is the case $p=2$, $q=1$.
We will show that the process $s[n] = u_1[n] + u_2[n]$ is an instance
of an ARMA(2,1) process and can thus be equivalently written as the
recurrence \cref{eq:SM-AR1} with coefficients
\begin{align}
  a_1 &= \phi_1 + \phi_2 \\
  a_2 &= -\phi_1 \phi_2 \\
  b_0 &= \frac{\theta_+ + \theta_-}{2} \label{eq:SM-b_0}\\
  b_1 &= \frac{\theta_- - \theta_+}{2} \label{eq:SM-b_1}\,.
\end{align}
In \cref{eq:SM-b_0,eq:SM-b_1} we used the shorthands
\begin{align}
  \theta_+^2 &= \theta_1^2(1+\phi_2)^2+ \theta_2^2(1 + \phi_1)^2 \\
  \theta_-^2 &= \theta_1^2(1-\phi_2)^2+ \theta_2^2(1 - \phi_1)^2 \,.
\end{align}

To derive these expressions we compare the power spectral density of
$u_1[n] + u_2[n]$ to the canonical power spectrum of an ARMA(2,1) process. The
AR(1) process $u_j[n]$ has the spectrum
$\theta_j^2/|1-\phi_j e^{-i\omega}|^2$. Since the PSDs of linear
processes are additive, the sum of both processes has the PSD
\begin{equation}
  S_s(\omega) = \frac{\theta_1^2}{|1-\phi_1 e^{-i\omega}|^2} +
  \frac{\theta_2^2}{|1-\phi_2 e^{-i\omega}|^2}
  = \frac{\theta_1^2 |1-\phi_2 e^{-i\omega}|^2 + \theta_2^2|1-\phi_1
  e^{-i\omega}|^2}{|1-\phi_1 e^{-i\omega}|^2\,|1-\phi_2 e^{-i\omega}|^2} \,.
  \label{eqn:SM-ar1sum_psd}
\end{equation}
The canonical spectrum of the ARMA(2,1) process defined by the
recurrence \cref{eq:SM-AR1} is
\begin{equation}
  S_{\rm ARMA}(\omega) = \frac{|b_0+b_1 e^{-i\omega}|^2}{|1 - a_1
  e^{-i\omega} - a_2 e^{-2i\omega}|^2} \,.
\end{equation}
Matching the denominators requires
$1 - a_1 z - a_2 z^2 = (1-\phi_1 z)(1-\phi_2 z)$ with $z=e^{-i\omega}$,
which yields $a_1 = \phi_1+\phi_2$ and $a_2 = -\phi_1\phi_2$.

Matching the numerators requires
\begin{equation}
  |b_0+b_1 e^{-i\omega}|^2 = \theta_1^2 |1-\phi_2 e^{-i\omega}|^2 +
  \theta_2^2|1-\phi_1 e^{-i\omega}|^2 \,.
  \label{eqn:SM-ar1sum_numerator_match}
\end{equation}
Both sides are linear combinations of $1$ and $\cos\omega$, so equality
for all $\omega$ follows from matching at any two values. Evaluating at
$\omega=0$ gives $(b_0+b_1)^2 = \theta_-^2$, and at $\omega=\pi$ gives
$(b_0-b_1)^2 = \theta_+^2$. Choosing positive roots fixes
$b_0+b_1=\theta_-$ and $b_0-b_1=\theta_+$, which yields the stated
values of $b_0$ and $b_1$.

\subsection{Mutual-Information Estimators and Training}
\label{sec:SM-nonlinear-training}

For each dataset we estimate $I(S_{1:T};X_{1:T})$ using ML-PWS, the
PWS ground-truth, the closed-form Gaussian baseline, the difference
of entropies estimator (DoE)~\cite{2020.McAllester}, and the contrastive
estimators MINE, SMILE~\cite{song2020understanding-e06}, and
InfoNCE~\cite{2018Oord}. The same estimator hyperparameters are used
across all $21$ values of the swept parameter in each panel.

\paragraph{DoE and ML-PWS.}
The DoE estimator decomposes $I = H(X_{1:T}) - H(X_{1:T}\mid
S_{1:T})$ and parameterizes the marginal $q_\theta(x_t\mid x_{1:t-1})$
and conditional $q_\theta(x_t\mid s_{1:t},x_{1:t-1})$ by two
autoregressive Gaussian recurrent networks, trained jointly to
minimize the negative log-likelihood
$-\mathbb{E}[\log q_\theta(x|s) + \log q_\theta(x)]$. Each network is
a stacked LSTM with $L=2$ layers and hidden width $h=32$, followed by
a linear head that outputs the per-step mean and log-variance of a
Gaussian; a zero-initialized linear skip
connection from the LSTM input directly to the head is used. The
conditional model
takes $(s_t, x_{t-1})$ as input; the marginal model
takes $x_{t-1}$ alone. A dropout of $0.05$ is applied between LSTM
layers. Both models are trained with Adam at learning rate $10^{-3}$
and batch size $64$ for up to $400$ epochs, with early stopping on
the validation loss; the checkpoint with the
lowest validation loss is retained.

To promote a comparison on equal footing,
ML-PWS reuses the trained DoE conditional model inside the RR-PWS
sequential Monte Carlo (SMC) scheme (\cref{sec:SM-marginalisation}) that
estimates the marginal
likelihood $q_\theta(x_{1:T})$ by averaging over input trajectories
sampled from the
true stimulus prior $p(s_{1:T})$ from \cref{eq:SM-AR1}. We use
$M=8192$ input trajectories per output sequence, processed in batches of
$64$ output sequences.

We apply conventional PWS to the stochastic model directly to obtain
the ground-truth result against which we compare the results of
ML-PWS and the other estimators.
For the ground-truth PWS estimate we use the same SMC settings as for
ML-PWS but replace
the learned conditional density with the analytic likelihood implied by
\cref{eq:SM-nonlinear_model}.

\paragraph{Gaussian baseline.}
Computed in closed form from the empirical joint covariance
$\widehat{\boldsymbol\Sigma}$ on the validation set.

\paragraph{Variational MI estimators (MINE, SMILE, InfoNCE).}
All three contrastive estimators share a bilinear
critic~\cite{song2020understanding-e06}: rather than parameterizing
the critic $g_\theta(s,x)$ as a single joint network, the score is
formed as the scaled inner product of two separately learned
embeddings of $s$ and $x$. Two neural-network encoders $g_S,g_X:
\mathbb{R}^T\to\mathbb{R}^{256}$, each consisting of two linear
layers of width $256$ separated by an ELU activation, are applied
to the flattened sequences $s_{1:T}$ and $x_{1:T}$ respectively. The
bilinear score function is
\begin{equation}
  T(s,x) = \frac{g_S(s)^\top g_X(x)}{\sqrt{256}}\,.
  \label{eq:SM-bilinear_score}
\end{equation}
This function is used as the critic $g_\theta(s,x)$ in
each variational objective. For a mini-batch of size $N$ this yields
an $N\times N$ score matrix
in which diagonal entries correspond to joint samples
$(s^{(i)},x^{(i)})$ and off-diagonal entries serve as
independently-sampled marginal pairs. Training uses Adam with
learning rate $10^{-3}$, weight decay $10^{-4}$, batch size $N=256$,
and gradient-norm clipping at $\|\nabla\|\le 1$, for $500$ epochs.
The three estimators differ only in the loss applied to the score
matrix: MINE uses the Donsker--Varadhan bound with an exponential
moving-average correction (decay $0.1$) to the partition-function
estimate; SMILE uses a clipped Donsker--Varadhan bound, with the
score clipped to $|T|\le\tau=5$, giving a value ceiling of
approximately $2\tau$; and InfoNCE uses a cross-entropy loss, whose
expectation is upper-bounded by $\log N \approx 5.54$ nats.

\section{Example III: Latent Factor Model}
\label{sec:SM-example3}

\label{sec:SM-vmm-benchmark}%
We use this model as a benchmark because it poses two distinct
challenges. First, the signal is encoded entirely in the output
variance rather than its mean as it commonly occurs in stochastic
volatility models in finance. Such a nonlinear dependence excludes
the use of Gaussian approximations. Second, the marginal
distribution of the two-dimensional
output is a heavy-tailed continuous mixture of Gaussians, which is more
difficult to learn than the conditional distribution. This is
particularly relevant for methods that train a separate neural network
for the marginal, such as DoE \cite{2020.McAllester}.
The benchmark model consists of two independent
Ornstein-Uhlenbeck processes driving a $d$-dimensional output
through a signal-modulated noise amplitude. The signal process
$S_t$ and a shared noise factor $Z_t$ evolve as
\begin{align}
  S_{t+1} &= e^{-1/\tau_s}\, S_t + \sqrt{1 - e^{-2/\tau_s}}\; \eta^s_t\,,
  \label{eqn:SM-ou_signal}\\
  Z_{t+1} &= e^{-1/\tau_z}\, Z_t + \sqrt{1 - e^{-2/\tau_z}}\; \eta^z_t\,,
  \label{eqn:SM-ou_noise}
\end{align}
where $\eta^s_t, \eta^z_t \sim \mathcal{N}(0,1)$ are independent.
The coefficients are chosen such that both processes have unit
stationary variance. Each output component $X^i_t$, $i=1,\ldots,d$,
is given by
\begin{equation}
  X^i_t = \ell\, e^{\alpha S_t}\, Z_t + \sigma\,
  \varepsilon^i_t\,,
  \label{eqn:SM-vmm_output}
\end{equation}
where $\varepsilon^i_t \sim \mathcal{N}(0,1)$ are independent
across components and time. The shared stochastic factor
$\ell\,e^{\alpha S_t}\,Z_t$ couples all output channels with an
amplitude that depends nonlinearly on the signal: the signal
modulates the variance and inter-channel covariance of the output
without affecting its conditional mean. Because $Z_t$ has zero mean,
$\mathbb{E}[X^i_t\mid S_t]=0$, $S$ and $X^i$
are linearly uncorrelated. A Gaussian estimator, which captures only
pairwise linear dependencies, would therefore find zero
mutual information despite there being a statistical dependence encoded in
the signal-modulated variance. The parameter values
used are listed in \cref{tab:SM-vmm_parameters}.

\begin{table}
  \centering
  \caption{Dataset parameters for the latent factor model
    [\cref{eqn:SM-ou_signal,eqn:SM-ou_noise,eqn:SM-vmm_output}] used in
    Fig.~1(d) of the main text. The training-set size
    $N_\mathrm{train}$ is varied while all other parameters are held
  fixed.}
  \label{tab:SM-vmm_parameters}
  \begin{tabular}{lll}
    \textbf{Parameter} & \textbf{Symbol} & \textbf{Value} \\
    \midrule
    \multicolumn{3}{l}{\textit{Output [\cref{eqn:SM-vmm_output}]}} \\
    Exponential gain & $\alpha$ & $1.0$ \\
    Factor loading & $\ell$ & $1.0$ \\
    Idiosyncratic noise std & $\sigma$ & $0.5$ \\
    Output dimension & $d$ & $2$ \\
    \midrule
    \multicolumn{3}{l}{\textit{Latent processes
    [\cref{eqn:SM-ou_signal,eqn:SM-ou_noise}]}} \\
    Stimulus timescale & $\tau_s$ & $50.0$ \\
    Stimulus stationary std & $\sigma_s$ & $1.0$ \\
    Latent-noise timescale & $\tau_z$ & $3.0$ \\
    \midrule
    \multicolumn{3}{l}{\textit{Dataset}} \\
    Sequence length & $T$ & $100$ \\
    Training trials & $N_\mathrm{train}$ & $\{20,\, 50,\, 100,\,
    200,\, 500,\, 1000,\, 2000,\, 5000\}$ \\
    Validation trials & $N_\mathrm{val}$ & $1000$ \\
  \end{tabular}
\end{table}

As for the nonlinear model of the previous subsection, it is useful
to relate the sequence
length to the input correlation time. The signal process $S_t$ has
correlation time $\tau_s=50$ (in units of the time step, cf.\
\cref{eqn:SM-ou_signal}), so each trajectory of length $T=100$ spans two
signal correlation times, $T/\tau_s = 2$. The effective
number of independent signal fluctuations in the training set is
therefore of order $N_\mathrm{train}\,T/\tau_s = 2\,N_\mathrm{train}$,
ranging from $\approx 40$ at $N_\mathrm{train}=20$ to $\approx 10^4$ at
$N_\mathrm{train}=5000$.

The marginal distribution of a single component $X^i$ is a
continuous mixture of Gaussians. Its tails are asymmetric:
for $S\ll 0$ the exponential factor $e^{\alpha S}\to 0$ and the
common input collapses, leaving an approximately Gaussian
left tail with scale $\sigma$; for $S\gg 0$ the shared input
dominates and $X^i \approx \ell e^{\alpha S} Z$, producing a
log-normal-type right tail. The marginal is therefore right-skewed.
This non-Gaussian character implies that estimators relying on a
Gaussian approximation to the marginal distribution will misrepresent
the true statistics and misestimate $I(S;X)$.

\subsection{Exact Likelihood}
\label{sec:SM-vmm-likelihood}
Although the marginal statistics of $X$ are non-Gaussian, the
conditional model
$\prob(X\mid S)$ itself is linear-Gaussian: given the
stimulus trajectory $S_{1:t}$, \cref{eqn:SM-vmm_output} is linear in
$Z_t$ with additive Gaussian noise. We can therefore evaluate the exact
sequential likelihood
$\ln \prob(x_t\mid x_{1:t-1}, s_{1:t})$ in closed
form by a scalar Kalman filter tracking $Z_t$. Writing the
signal-modulation as $a_t \equiv \ell\,e^{\alpha S_t}$, the
one-step predictive distribution of the $d$-dimensional output is the
Gaussian
\begin{equation}
  \prob(X_t \mid X_{1:t-1}, S_{1:t}) =
  \frac{\exp\!\left[-\tfrac{1}{2}\,
      (X_t - \boldsymbol\mu_t)^\top \boldsymbol\Sigma_t^{-1}
  (X_t - \boldsymbol\mu_t)\right]}
  {\sqrt{(2\pi)^d \det \boldsymbol\Sigma_t}}\,,
  \label{eqn:SM-vmm_innovation}
\end{equation}
with mean and covariance
\begin{equation}
  \boldsymbol\mu_t = a_t\, m_t\, \mathbf{1}\,,\qquad
  \boldsymbol\Sigma_t = a_t^2\, v_t\, \mathbf{1}\mathbf{1}^\top +
  \sigma^2\, \mathbf{I}_d\,,
  \label{eqn:SM-vmm_moments}
\end{equation}
where $m_t = \mathbb{E}[Z_t\mid X_{1:t-1}, S_{1:t}]$ and $v_t =
\operatorname{Var}[Z_t\mid X_{1:t-1}, S_{1:t}]$ are the predictive mean
and variance of $Z_t$,
$\mathbf{1}=(1,\ldots,1)^\top\in\mathbb{R}^d$, and $\mathbf{I}_d$ is
the identity.
The pair $(m_t, v_t)$ is propagated by the standard Kalman
recursion~\cite{1960.Kalman}, initialized at the stationary values
$m_1 = 0$ and $v_1 = 1$ of \cref{eqn:SM-ou_noise}. The covariance
$\boldsymbol\Sigma_t$ is
the sum of a rank-one matrix and a diagonal matrix:
a single shared factor correlating all
$d$ channels, plus per-channel noise of variance
$\sigma^2$; its inverse and determinant, needed to evaluate the
Gaussian density, follow from the Sherman--Morrison
formula~\cite{1950.Sherman}. As in the nonlinear model
(\cref{sec:SM-example2}), the ground-truth PWS estimate is obtained
by supplying this analytic likelihood to the RR-PWS sequential Monte
Carlo scheme (\cref{sec:SM-marginalisation}), so it is unbiased and does
not depend on any learned model or on the training set size
$N_\mathrm{train}$.

\subsection{Learned Models: Diagonal vs.\ Low-Rank Covariance}
\label{sec:SM-vmm-rank}
Unlike the ground-truth PWS estimate, which uses the exact Kalman
conditional of \cref{sec:SM-vmm-likelihood}
[\cref{eqn:SM-vmm_innovation,eqn:SM-vmm_moments}], the two learned estimators
DoE and ML-PWS both replace that exact conditional by a \emph{learned}
autoregressive Gaussian model, as in the nonlinear-model benchmark.
Concretely, the conditional density is represented by an LSTM that
reads, at each step $t$, the history and the current stimulus $s_t$,
and updates a hidden state $h_t$ summarizing the history seen so far. A
linear head then maps $h_t$ to the parameters of a Gaussian predictive
distribution over the $d$-dimensional output $x_t$---a mean vector
$\boldsymbol\mu_t\in\mathbb{R}^d$ together with the parameters of the
covariance $\boldsymbol\Sigma_t$---defining the conditional density
$q_\theta(x_t\mid x_{1:t-1},s_{1:t}) =
\mathcal{N}(x_t;\boldsymbol\mu_t,\boldsymbol\Sigma_t)$.

The two learned estimators differ only in \emph{which} densities they
learn, not in how the conditional is parameterized. ML-PWS learns
\emph{only} this conditional and obtains the marginal by exact
marginalization over the known stimulus prior $p(s_{1:T})$; it does
\emph{not} train a marginal network.
DoE learns the same conditional model, and, in addition, trains a
separate, identically structured network that omits the stimulus input
to represent the marginal density $q_\theta(x_t\mid x_{1:t-1})$.

The consequence for the two estimates is direct: DoE forms
$I = H(X_{1:T}) - H(X_{1:T}\mid S_{1:T})$ from the two fitted densities
$Q_\theta(x_{1:T})$ and $Q_\theta(x_{1:T}|s_{1:T})$, whereas ML-PWS uses
only the learned conditional $Q_\theta(x_{1:T}|s_{1:T})$ and obtains the
marginal likelihood $Q(x_{1:T})$ by averaging that conditional over
stimulus trajectories drawn from the exact prior $p(s_{1:T})$.

The covariance family and hence the number of parameters the neural
network model
must emit per step is set by a rank hyperparameter $r$:
\begin{itemize}
  \item \textbf{Diagonal ($r=0$).} $\boldsymbol\Sigma =
    \operatorname{diag}(\mathbf{d})$, i.e. a Multivariate Normal
    distribution that treats the $d$ output components
    as conditionally independent. Here the head predicts only the $d$
    diagonal (log-)variances $\mathbf{d}$ per step, in addition to
    the mean. This family \emph{cannot} represent
    the cross-channel correlation $\langle X_i X_j \rangle$ in
    \cref{eqn:SM-vmm_moments} and is therefore misspecified.
  \item \textbf{Rank-1 ($r=1$).} $\boldsymbol\Sigma =
    \mathbf{w}\mathbf{w}^\top + \operatorname{diag}(\mathbf{d})$, a
    Multivariate Normal density with rank-1 correlations. The head
    now predicts both the diagonal $\mathbf{d}$ and an additional
    loading vector $\mathbf{w}\in\mathbb{R}^d$ per step, and this
    single vector $\mathbf{w}$ captures the cross-channel covariance
    structure. This covariance can exactly represent
    \cref{eqn:SM-vmm_moments}, so this model family is
    correctly specified.
\end{itemize}
We run the entire calculation under both modeling assumptions to
understand the effect of covariance misspecification.

We emphasize that ``correctly specified'' here refers to the
\emph{conditional} model only. The true conditional density is 2D-Gaussian
[\cref{eqn:SM-vmm_innovation}], so a Gaussian model with a rank-1 covariance
can represent it exactly. The true marginal density, in contrast, is a
continuous mixture of Gaussians and is heavy-tailed, so
a Gaussian marginal model is misspecified. This
asymmetry is precisely what separates the two estimators: ML-PWS uses
only the conditional model, and marginalizes it exactly over the known
stimulus prior, whereas DoE additionally relies on the Gaussian marginal
model, whose irreducible misspecification biases its estimate of
$H(X_{1:T})$ and hence of the mutual information.

\subsection{Training and Estimation Hyperparameters}
\label{sec:SM-vmm-training}
The DoE, ML-PWS, and PWS estimators use the same architecture and
optimizer as in the nonlinear-model benchmark
(\cref{sec:SM-nonlinear-training}), with the following settings. Each Gaussian
recurrent network is a stacked LSTM with $L=2$ layers, hidden width
$h=64$, and dropout $0.05$ between layers, followed by the
rank-$r$ Gaussian head of \cref{sec:SM-vmm-rank}. Both the conditional
and marginal networks are trained jointly with Adam at learning rate
$10^{-3}$ and batch size $64$ for up to $400$ epochs. ML-PWS and PWS use the
RR-PWS SMC scheme with $M=1024$ input trajectories per output
sequence.

The contrastive estimators (MINE, SMILE, InfoNCE) use the bilinear
critic of \cref{eq:SM-bilinear_score} with two layers (width $256$),
trained with Adam at learning rate $10^{-3}$, gradient-norm
clipping $\|\nabla\|\le 1$, and batch size $N=256$ for $500$ epochs;
SMILE clips the score at $\tau=5$. Each contrastive estimate is
averaged over $3$ random seeds.

\subsection{Benchmark Outcome}
\label{sec:SM-vmm-results}
Figure 1d of the main text reports the final-step ($t=T=100$) mutual
information while varying the training data size $N_\mathrm{train}$.
The PWS ground truth does not depend on the amount of training data,
since it is computed from the exact analytic likelihood
(\cref{sec:SM-vmm-likelihood}); it is $\approx 10.0$~nats.
ML-PWS converges upward toward this reference as
$N_\mathrm{train}$ grows: in the correctly-specified rank-1 regime
it reaches the truth ($9.93$~nats at $N_\mathrm{train}=5000$), whereas
in the misspecified diagonal (rank-0) regime it saturates near $8.3$~nats
because a diagonal covariance structurally cannot capture the
shared-factor correlation. DoE significantly over-estimates the mutual
information ($13$~nats in the best case), and is worse at small
$N_\mathrm{train}$, since
its learned Gaussian marginal model is insufficiently expressive to
capture the heavy-tailed distribution of the data;
ML-PWS avoids this because its marginal is
obtained from the exact stimulus prior via RR-PWS rather than from a
separate marginal model.
The variational neural network estimators fail outright: the true MI
($\approx 10$~nats) sits at or above the InfoNCE ceiling
$\log N = \log 256 \approx 5.54$~nats and the SMILE ceiling
$2\tau = 10$~nats, and MINE diverges.

\section{Neural Ganglion Cells}
\label{sec:SM-ganglion}

\subsection{Dataset}

We analyze recordings of \num{230} retinal ganglion cells from a
salamander, captured as spike trains at a native temporal
resolution of \SI{0.1}{\milli\second}.
The full recording has a
duration of approximately \SI{8148}{\second} and contains two types of
segments.
\emph{Non-repeating} segments present a continuously novel,
stochastically generated stimulus; these provide diverse
stimulus--response pairs for training the neural network, and a
held-out portion (never seen during training) is used to obtain
the PWS mutual information estimate.
\emph{Repeating} segments consist of \num{136} presentations of an
identical stimulus of duration $T=\SI{30.08}{\second}$; because the
same stimulus is replayed across trials, the trial-to-trial variability
of the output that arises from the intrinsic noise of the system can be
separated from the output variability that arises from the propagation
of the input variability to the output. These repeating segments
provide the validation data.

\begin{figure}
  \centering
  \includegraphics[width=0.6\linewidth]{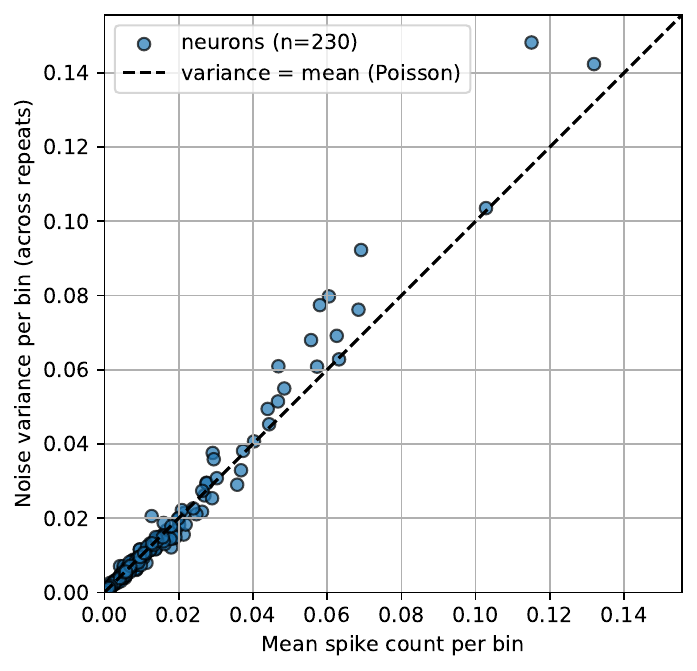}
  \caption{Spike-count statistics are consistent with a Poisson
    model. For each of the 230 neurons, spike counts were computed in
    10 ms bins across the 136 repeated presentations of the identical
    validation stimulus. Each point shows the across-repeat noise
    variance $\sigma^2_t$ versus the mean spike count per bin
    $\mu_t$, averaged over time.
    Points fall close to the identity line,
    as expected for Poisson firing (median Fano factor
  $\sigma^2_t/\mu_t = 0.98$).}
  \label{fig:SM-poisson_assumption}
\end{figure}

We discretize the time-series data with a resolution of $\Delta t
= \SI{10}{\milli\second}$. The spike count of neuron $k$ in the bin
$[t_i, t_{i+1})$ is denoted $x^k_i$ and modeled as a Poisson random
variable with intensity $\lambda^k_i$. The Poisson assumption has
been empirically tested, see \cref{fig:SM-poisson_assumption}.
The stimulus is rescaled so that its variance over the recording is
$\langle s^2\rangle=1$.
For training we partition the non-repeating segments into windows of
length \SI{4}{\second} with \SI{50}{\percent} overlap, yielding
\num{521} non-repeating \SI{4}{\second} windows used for training (the
same \num{521} samples swept in \cref{fig:SM-data_study}); the
validation data, used to validate the model as described in
\cref{sec:SM-ganglion-validation} below, are the \num{136} repeated trials
taken in their entirety.
Note that, because of the \SI{50}{\percent} overlap, adjacent windows
are not statistically independent.

\subsection{Stimulus}

The stimulus is the position $s(t)$ of a vertically
moving bar, drawn from a stochastically driven harmonic oscillator
(SHO)~\cite{Marre.2015,sederberg2018learning-3e3,sachdeva2021optimal-f51}.
The SHO is described by a second-order linear stochastic differential equation,
\begin{align}
  \frac{\mathrm{d}s}{\mathrm{d}t} &= v\,, \label{eqn:SM-sho_x}\\
  \frac{\mathrm{d}v}{\mathrm{d}t} &= -\omega_0^2\,s -
  \frac{v}{\tau} + \omega_0\sqrt{\frac{2}{\tau}}\,\xi(t)\,, \label{eqn:SM-sho_v}
\end{align}
where $\xi(t)$ is unit-strength Gaussian white noise,
$\tau=\SI{50}{\milli\second}$ is the relaxation time, and
$\omega_0=\SI{9.42}{\radian\per\second}$ corresponds to the
resonance frequency $f_0=\omega_0/(2\pi)\approx\SI{1.5}{\hertz}$.
With these values the dimensionless damping coefficient
$\gamma=1/(2\omega_0\tau)\approx\num{1.06}$, so the oscillator is
slightly over-damped. The diffusion strength is chosen so that the
stationary variance of $s$ equals one.

In state-space form, $\mathbf{y}=(s,v)^\top$ obeys the linear It\^o
SDE $\mathrm{d}\mathbf{y} = A\,\mathbf{y}\,\mathrm{d}t +
B\,\mathrm{d}W_t$ with
\begin{equation}
  A =
  \begin{pmatrix} 0 & 1 \\ -\omega_0^2 & -\tau^{-1}
  \end{pmatrix}\,,\qquad B =
  \begin{pmatrix} 0 \\ \sqrt{2D}/\tau
  \end{pmatrix}\,,
  \label{eqn:SM-sho_state_space}
\end{equation}
with $D=\omega_0^2\tau$. Because the drift is linear, the process
admits an exact discrete-time propagator $\Phi=\exp(A\,\Delta t)$,
i.e., the recursion
$\mathbf{y}_{i+1}=\Phi\,\mathbf{y}_i+\boldsymbol{\eta}_i$ is
statistically equivalent to the continuous process being sampled at discrete
times. The eigenvalues of $\Phi$ are the
poles of the corresponding discrete transfer function and
characterize the system's relaxation timescales: for $\Delta t =
\SI{10}{\milli\second}$ both poles are real ($z_{1,2}\approx
0.936,\,0.875$), reflecting two decaying exponential modes
in the over-damped regime. The discrete-time stimulus
is therefore an AR(2) process.

Crucially, because $P(s)$ is fully specified by the known AR(2)
parameters, ML-PWS can draw synthetic stimulus trajectories directly
from this prior to marginalize the learned joint distribution
$P(s,x)$ via the RR-PWS scheme.

\subsection{Spike Model}
\label{sec:SM-ganglion-spike}

\begin{figure*}
  \includegraphics[width=\textwidth]{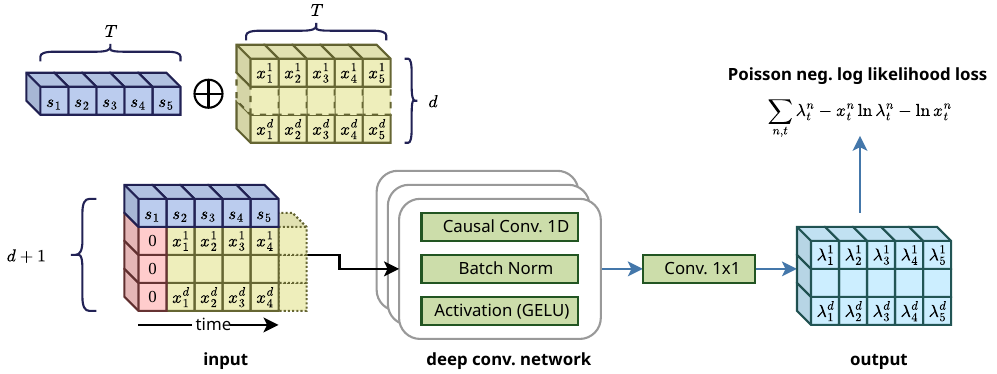}
  \caption{Neural network architecture of deep convolutional model
    for neuronal activations. We combine the input time-series
    $s_{1:T}=s_1,\ldots,s_T$ with the $d$-dimensional output time
    series $\boldsymbol{x}_{1:T} = (x^1_{1:T},\ldots,x^d_{1:T})^\top$
    where $d$ is the number of neurons into a $(d+1)\times T$ matrix.
    We feed this into the causal 1D convolutional neural network.
    The outputs are shifted right such that the network learns to
    predict the next time step. The results
    are 1-step predictions for the conditional Poisson intensities
    for each neuron. The network is trained with the Poisson negative
  log likelihood loss.}
  \label{fig:SM-cnn_cartoon}
\end{figure*}

We model the conditional spike-count distribution
$\prob(x^k_{1:n}|s_{1:n})$ with a one-dimensional causal
convolutional neural network, see \cref{fig:SM-cnn_cartoon}. The
stimulus $s(t)$ is processed by a
stack of 1D convolutional layers, each followed by batch
normalization and a ReLU activation. Causality is enforced by
left-padding the input so that the receptive field at time $t_i$
extends only to past stimulus values $s_{j\leq i}$. A final
$1\times 1$ convolution maps the feature stack to per-neuron
log-intensities $\ln\lambda^k_i$, and the spike counts $x^k_i$ are
taken to be conditionally independent Poisson random variables with
mean $\lambda^k_i\,\Delta t$.
For training the collective model, the negative log-likelihood
minimized during training is
\begin{equation}
  \mathcal{L} = \sum_{i,k}\Bigl[ -x^k_i\ln(\lambda^k_i\Delta t) +
  \lambda^k_i\Delta t + \ln(x^k_i!)\Bigr],
  \label{eqn:SM-poisson_nll}
\end{equation}
where the sum runs over all time bins $i$ and all neurons $k$
simultaneously. The loss for an individual-neuron model is identical
but with the sum restricted to a single $k$, fitting one network per
neuron independently.
We restrict the analysis to a subset of \num{50} of the \num{230}
recorded ganglion cells; this is a proof-of-concept choice rather than
a fundamental limit of the method. We train two distinct types of
model, each built from four causal convolutional layers with kernel
size \num{8}. The \emph{single-neuron} models are \num{50} independent
networks, one trained per neuron, each with a hidden size of \num{20}
channels and a single output. The \emph{collective} model is one
network with a hidden size of \num{256} channels whose \num{50} output
channels cover all selected cells simultaneously.

Although trained purely as a one-step-ahead predictor, the
model is also a full generative model of spiking. Because the
network is autoregressive---each prediction is conditioned on the
stimulus and on past spike counts---it can be run in
\emph{generative mode} to synthesize novel spike trajectories for
any input stimulus $s_{1:T}$: starting from an empty history, at
each bin $i$ we evaluate the intensities $\lambda_i^k$, draw
$\tilde x_i^k\sim\mathrm{Poisson}(\lambda_i^k\,\Delta t)$, append
the sampled counts to the running history, and advance to bin
$i+1$. Iterating to $i=T$ yields a complete sampled response
$\tilde x_{1:T}^k$; repeating the procedure produces
independent draws from the model's conditional distribution
$\prob(x_{1:T}|s_{1:T})$. We make use of this generative ability
for model validation and for the channel capacity computation,
as described below.

\subsection{Training}
\label{sec:SM-ganglion-training}

The network is trained on the partitioned non-repeating block, which
provides $N_\text{train}=\num{521}$ stimulus--response trajectory pairs
of duration \SI{4}{\second}, with a batch size of \num{32}. We
minimize the Poisson NLL
[\cref{eqn:SM-poisson_nll}] using the Adam
optimizer~\cite{2014.Kingma}. Training proceeds for \num{50}
epochs and the model with the lowest validation loss is
retained for the downstream PWS computation.

\subsection{Model Validation}
\label{sec:SM-ganglion-validation}

\begin{figure}[!ht]
  \centering
  \includegraphics[width=0.9\linewidth]{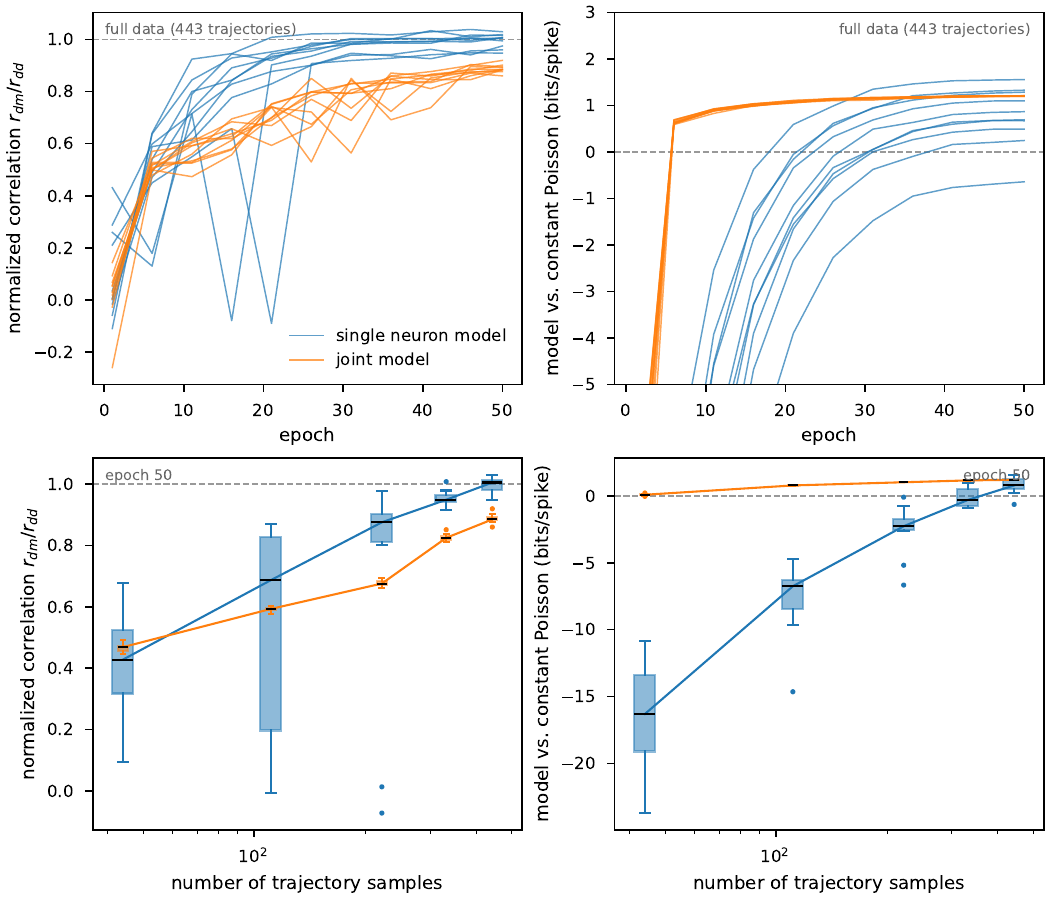}
  \caption{Effect of training-set size $N_\text{train}$ on predictive
    performance.
    A single-neuron CNN (blue) and the collective model of all 50
    neurons (orange) were each trained 10 times on increasing fractions
    of the 521 available 4~s non-repeating trajectory samples, with
    independent data draws and weight initializations.
    Performance is quantified on the held-out stimulus-repeat trials
    by two metrics.
    (Left column) Normalized PSTH correlation $r_{dm}/r_{dd}$,
    where $r_{dm}$ is the model-data correlation and
    $r_{dd}$ the data–data correlation (noise ceiling).
    (Right column) Log-likelihood gain over a per-neuron
    constant-rate Poisson baseline, in bits per spike;
    dashed line at 0 marks the baseline.
    (Top row) Per-epoch validation curves over 50 training epochs for
    the full-data runs; each curve is one training run for a single
    representative neuron.
    (Bottom row) Final-epoch performance vs.\ the training-set size
    $N_\text{train}$, i.e.\ the number of stimulus--response trajectory
    pairs used to train the model;
  points are means over the 10 runs.}
  \label{fig:SM-data_study}
\end{figure}

We evaluate each trained model on the held-out repetition trials using
two complementary metrics: a \emph{generative} metric based on the
PSTH correlation, and a \emph{likelihood-based} metric in bits per
spike (\cref{fig:SM-data_study}).

\paragraph{Setup and held-out data.}
Validation uses the $B=136$ stimulus-repetition trials, yielding
spike counts $x_i^{k,(b)}$, where $k$ indexes the neuron, $i$ the
time bin, and $b=1,\dots,B$ the trial. The trial-averaged
spike-count profile
\begin{equation}
  \bar d_i^k = \tfrac{1}{B}\sum_{b=1}^B x_i^{k,(b)}
\end{equation}
is the peri-stimulus time histogram (PSTH): the deterministic,
stimulus-locked component of each neuron's response, with
trial-to-trial variability averaged out.
Both the single-neuron model and the collective model produce
per-neuron predicted intensities $\lambda_i^k$: for the single-neuron
model, $\lambda_i^k$ is the output of the network trained on neuron
$k$ alone; for the collective model it is the $k$-th output channel
of the joint network. The following metrics are computed for each
model type separately, yielding a per-neuron score in both cases.

\paragraph{Generative metric: PSTH correlation.}
Each model is run in generative mode: conditioned on the
repeat-trial stimulus, it draws $\tilde
x_i^k\sim\mathrm{Poisson}(\lambda_i^k\,\Delta t)$
at each bin and feeds its own samples back as history. Averaging over
$B$ generated trajectories gives the model PSTH, i.e.\ the model
prediction for the mean number of spikes of neuron $k$ at time point
$i$, $\bar m_i^k=\tfrac1B\sum_b \tilde x_i^{k,(b)}$.
The per-neuron correlation between the model prediction $\bar m_i^k$
and the data $\bar d_i^k$,
\begin{equation}
  r_{dm}^{k} = \mathrm{corr}_i\!\left(\bar m_i^k,\,\bar d_i^k\right)\,,
\end{equation}
is normalized by the split-half data--data correlation
\begin{equation}
  r_{dd}^{k}
  = \mathrm{corr}_i\!\left(\bar d_i^{k,\,\mathrm{half\,1}},\,
  \bar d_i^{k,\,\mathrm{half\,2}}\right),
\end{equation}
which sets the noise ceiling imposed by trial-to-trial variability
arising from the intrinsic noise in the system.
We report the mean normalized correlation $r_{dm}/r_{dd}$, with $1$
indicating that the model reaches the maximal value obtainable from
the data. Values
above $1$ are possible: the noise ceiling $r_{dd}$ is based on
trial-averaged PSTHs and does not account for history-dependent or
nonlinear temporal structure; a model that captures such structure
can predict the true underlying rate more accurately than what the
split-half data comparison reflects.

\paragraph{Likelihood metric: bits per spike.}
Each model is evaluated directly on the validation data: $\lambda_i^{k,(b)}$
is the Poisson rate predicted by the trained network when conditioned
on the true observed spike history of trial $b$ up to time bin $i$
(rather than on its own generated samples), giving the exact
log-likelihood
\begin{equation}
  \log q_\theta\!\left(x \mid s\right)
  = \sum_{k,i,b}
  \Big[\,x_i^{k,(b)}\log\!\big(\lambda_i^{k,(b)}\Delta t\big)
    - \lambda_i^{k,(b)}\Delta t
  - \log\!\big(x_i^{k,(b)}!\big)\Big].
\end{equation}
Model quality is expressed as the log-likelihood gain over a
homogeneous (constant-rate) Poisson null with per-neuron rates
$\lambda_0^k = \tfrac{1}{TB}\sum_{i,b} x_i^{k,(b)}$, normalized
per spike:
\begin{equation}
  \Lambda = \frac{\log q_\theta(x\mid s) - \log q_{0}(x)}{S\,\ln 2}
  \quad \text{[bits per spike]},
  \qquad S = \sum_{k,i,b} x_i^{k,(b)}.
\end{equation}
Positive $\Lambda$ indicates the model captures structure beyond the mean
firing rate.

The results are shown in \cref{fig:SM-data_study} as a function of
$N_\text{train}$, and as a function of training epochs. Both
metrics improve monotonically with $N_\text{train}$ and the training
curves converge within 50 epochs on the full dataset.

\begin{longtable}{SSSSSSS}
  \caption{Validation metrics per neuron, ordered by the neurons'
    firing rates. The firing rate and the split-half data-data
    correlation $r_\text{dd}$ are properties of the validation data;
    $r_\text{dm}$ is the model-data correlation and bits per spike the
    validation log-likelihood gain over a constant-rate Poisson
    process, each given for the single-neuron and the collective model.
    The correlations $r_\text{dd}$ and $r_\text{dm}$ are the raw
    Pearson correlations; the normalized metric
    $r_\text{dm}/r_\text{dd}$ plotted in \cref{fig:SM-data_study} is their
  ratio.} \label{tab:SM-neurons} \\
  \toprule
  {neuron index} & {firing rate} & {$r_\text{dd}$} &
  \multicolumn{2}{c}{$r_\text{dm}$} & \multicolumn{2}{c}{bits per spike} \\
  \cmidrule(lr){4-5} \cmidrule(lr){6-7}
  {} & {(spikes/s)} & {} & {single} & {collective} & {single} & {collective} \\
  \midrule
  \endfirsthead
  \caption[]{Validation metrics per neuron} \\
  \toprule
  {neuron} & {firing rate} & {$r_\text{dd}$} &
  \multicolumn{2}{c}{$r_\text{dm}$} & \multicolumn{2}{c}{bits per spike} \\
  \cmidrule(lr){4-5} \cmidrule(lr){6-7}
  {} & {(spikes/s)} & {} & {single} & {collective} & {single} & {collective} \\
  \midrule
  \endhead
  \midrule
  \multicolumn{7}{r}{Continued on next page} \\
  \midrule
  \endfoot
  \bottomrule
  \endlastfoot
  226 & 13.19 & 0.738 & 0.746 & 0.775 & 0.240 & 0.286 \\
  220 & 11.51 & 0.716 & 0.710 & 0.763 & 0.481 & 0.515 \\
  219 & 10.28 & 0.045 & 0.075 & 0.092 & 0.120 & 0.120 \\
  201 & 6.91 & 0.670 & 0.723 & 0.749 & 1.156 & 1.138 \\
  223 & 6.84 & 0.571 & 0.507 & 0.549 & 0.272 & 0.353 \\
  215 & 6.32 & 0.146 & 0.162 & 0.145 & 0.054 & 0.114 \\
  225 & 6.26 & 0.706 & 0.519 & 0.514 & 0.695 & 0.778 \\
  196 & 6.05 & 0.683 & 0.593 & 0.635 & 0.404 & 0.387 \\
  214 & 5.80 & 0.776 & 0.812 & 0.814 & 0.935 & 1.054 \\
  172 & 5.73 & 0.697 & 0.504 & 0.535 & 0.429 & 0.654 \\
  221 & 5.57 & 0.559 & 0.237 & 0.341 & 0.303 & 0.315 \\
  218 & 4.84 & 0.348 & 0.421 & 0.452 & 0.206 & 0.291 \\
  209 & 4.69 & 0.633 & 0.572 & 0.609 & 0.856 & 0.700 \\
  222 & 4.67 & 0.385 & 0.410 & 0.424 & 0.232 & 0.310 \\
  208 & 4.44 & 0.661 & 0.674 & 0.708 & 0.786 & 0.855 \\
  217 & 4.40 & 0.792 & 0.728 & 0.755 & 1.076 & 1.295 \\
  227 & 4.03 & 0.358 & 0.392 & 0.432 & 0.082 & 0.135 \\
  155 & 3.73 & 0.599 & 0.640 & 0.580 & 1.203 & 1.062 \\
  123 & 3.68 & 0.792 & 0.864 & 0.827 & 2.082 & 2.083 \\
  110 & 3.58 & 0.702 & 0.688 & 0.675 & 1.510 & 1.166 \\
  228 & 3.03 & 0.182 & 0.224 & 0.187 & 0.053 & 0.073 \\
  211 & 2.94 & 0.492 & 0.238 & 0.310 & 0.790 & 0.768 \\
  135 & 2.92 & 0.702 & 0.798 & 0.794 & 1.238 & 1.167 \\
  195 & 2.90 & 0.834 & 0.655 & 0.650 & 1.257 & 1.577 \\
  171 & 2.75 & 0.387 & 0.545 & 0.423 & 0.213 & 0.233 \\
  76 & 2.75 & 0.508 & 0.645 & 0.620 & 1.006 & 1.198 \\
  165 & 2.69 & 0.777 & 0.804 & 0.812 & 1.778 & 1.768 \\
  210 & 2.63 & 0.547 & 0.519 & 0.541 & 0.595 & 0.811 \\
  45 & 2.63 & 0.642 & 0.738 & 0.681 & 1.823 & 1.322 \\
  111 & 2.48 & 0.716 & 0.498 & 0.516 & 0.908 & 1.108 \\
  204 & 2.39 & 0.762 & 0.618 & 0.585 & 1.445 & 1.765 \\
  199 & 2.19 & 0.725 & 0.637 & 0.604 & 1.249 & 1.468 \\
  24 & 2.15 & 0.565 & 0.676 & 0.620 & 1.584 & 1.506 \\
  100 & 2.13 & 0.840 & 0.663 & 0.680 & 2.252 & 2.508 \\
  193 & 2.08 & 0.552 & 0.399 & 0.406 & 0.671 & 0.892 \\
  83 & 2.02 & 0.441 & 0.456 & 0.453 & 0.527 & 0.254 \\
  169 & 2.01 & 0.670 & 0.502 & 0.549 & 0.793 & 1.244 \\
  20 & 2.01 & 0.596 & 0.645 & 0.625 & 1.296 & 1.444 \\
  207 & 1.98 & 0.813 & 0.787 & 0.827 & 2.400 & 2.497 \\
  185 & 1.91 & 0.828 & 0.774 & 0.755 & 2.223 & 2.615 \\
  152 & 1.91 & 0.686 & 0.539 & 0.479 & 1.147 & 1.398 \\
  102 & 1.86 & 0.744 & 0.760 & 0.758 & 1.996 & 2.249 \\
  80 & 1.84 & 0.702 & 0.669 & 0.601 & 1.123 & 1.335 \\
  213 & 1.80 & 0.709 & 0.493 & 0.459 & 0.754 & 1.025 \\
  4 & 1.80 & 0.782 & 0.502 & 0.516 & 2.244 & 2.361 \\
  120 & 1.69 & 0.728 & 0.520 & 0.500 & 0.900 & 1.197 \\
  78 & 1.67 & 0.783 & 0.747 & 0.760 & 1.626 & 1.347 \\
  157 & 1.60 & 0.583 & 0.418 & 0.551 & 1.360 & 1.719 \\
  87 & 1.57 & 0.074 & 0.046 & 0.098 & 0.485 & -0.055 \\
  216 & 1.57 & 0.858 & 0.720 & 0.795 & 2.261 & 2.840 \\
  \midrule
  {mean} & 3.77 & 0.616 & 0.564 & 0.571 & 1.022 & 1.105 \\
\end{longtable}

\subsection{From Mutual Information to Information
Rates}\label{sec:SM-ganglion_rates}

ML-PWS returns the path mutual information $\hat{I}(T)$ as a
function of the trajectory duration $T=n\Delta t$ (the bias-corrected
  estimate obtained from the $1/M\to0$ extrapolation of
\cref{sec:SM-bias}; see \cref{fig:SM-marginal_bias}). The information rate
is the asymptotic slope of this curve, $R=\lim_{T\to\infty}\hat{I}(T)/T$,
which for a stationary process is also the slope of the linear regime of
$\hat{I}(T)$. We therefore estimate it as a secant slope between the
longest duration available, $T=\SI{4}{\second}$, and a
reference time $\tau_0=\SI{0.5}{\second}$,
\begin{equation}
  \hat{R} =
  \frac{\hat{I}(T) - \hat{I}(\tau_0)}
  {T - \tau_0}\,.
  \label{eqn:SM-rate_secant}
\end{equation}
The single-neuron and
collective rates reported in Fig.~2(c) of the main text are both
obtained in this way, the former from the $50$ individual models and the
latter from the collective model.

We discard the initial part of the curve for $T<\tau_0$, since
$\hat{I}(T)$ is not necessarily linear there. The response at the
beginning of a
trajectory reflects the stimulus that preceded it, and this
onset contributes an offset to $\hat{I}(T)$.
Choosing $\tau_0 = \SI{0.5}{\second}$ greater than the correlation
time of the neurons
ensures that \cref{eqn:SM-rate_secant}  measures the asymptotic growth.

\subsection{Channel Capacity Optimization}
\label{sec:SM-capacity}

\begin{figure}
  \includegraphics[width=0.7\linewidth]{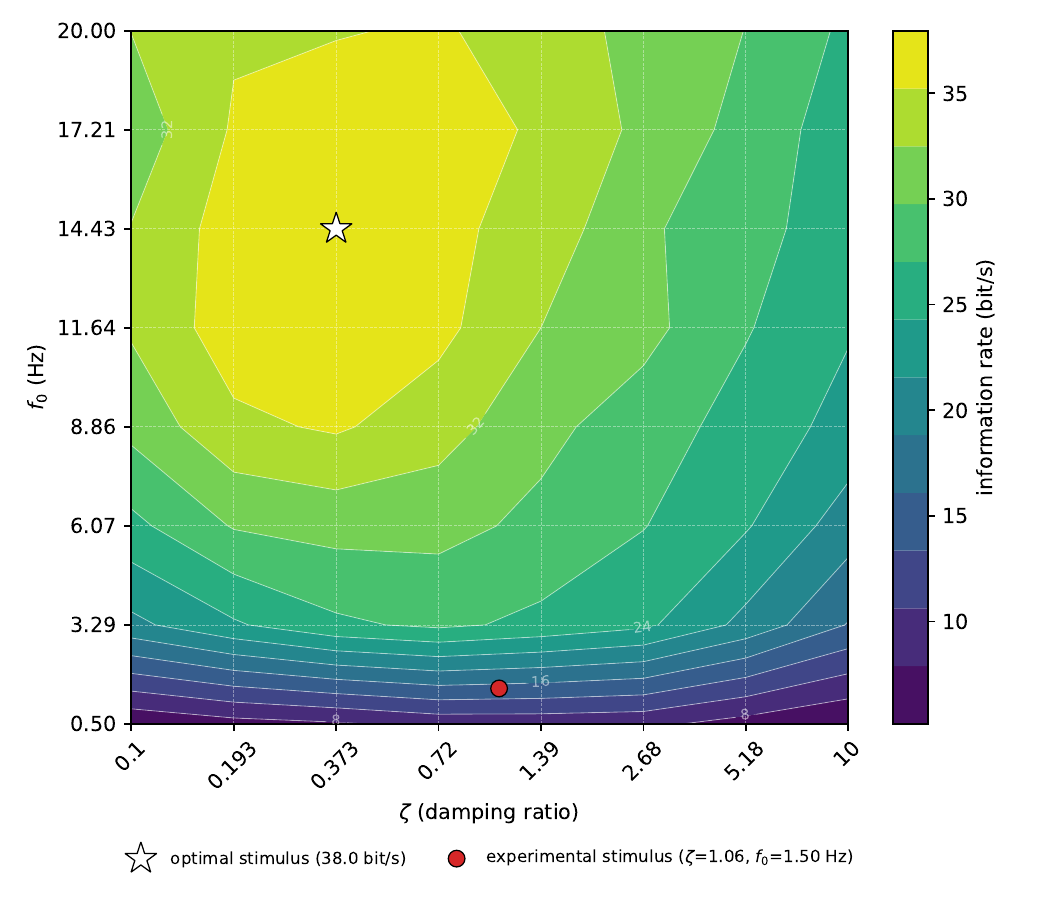}
  \caption{\textbf{Mutual-information landscape over the input distribution.}
    Grid-search of $I(S_{1:T};X_{1:T})$ over the two parameters
    specifying the input distribution: $8\times8$ points, $\zeta$
    logarithmically spaced over
    $[0.1,10]$ and $f_0$ linearly spaced over $[0.5,20]\,\mathrm{Hz}$. At each
    point ML-PWS estimates the mutual information ($M=4096$ RR-PWS trajectories,
    $0.5\,\mathrm{s}$ warm-up); colour is the information rate.
    White star: optimal stimulus.
    Red circle: the experimental stimulus
  ($\zeta=1.06$, $f_0=1.50\,\mathrm{Hz}$), well below the optimum.}
\end{figure}

Once the spike model has been trained, we estimate the channel
capacity of the population code by maximizing the mutual information
$I(S_{1:T};X_{1:T})$ over an input distribution $P_\theta(s_{1:T})$
while holding the channel $P(x|s)$ defined by the trained CNN
fixed. The optimization is constrained to inputs of unit power,
$\mathbb{E}[s_t^2]=1$.

\paragraph{Stochastic harmonic oscillator input family.}
We restrict the search to the same AR(2) family used to model the
recorded stimulus: the stationary Gaussian process obtained by exact
discretization of the stochastic harmonic oscillator (SHO)
\begin{equation}
  \ddot{s} + 2\zeta\omega_0\,\dot{s} + \omega_0^2\,s = \varepsilon(t)\,,
  \label{eqn:SM-sho_sde}
\end{equation}
parameterized by its natural frequency $\omega_0=2\pi f_0$ and damping
ratio $\zeta>0$.
Matched-pole discretization at the sampling interval
$\Delta t$ maps the continuous-time poles to discrete poles
\begin{equation}
  z_{1,2} = e^{\,\omega_0(-\zeta\pm\sqrt{\zeta^2-1})\,\Delta t}.
\end{equation}
From these poles, we obtain the corresponding AR(2) recurrence
\begin{equation}
  s_t = c_1\,s_{t-1} + c_2\,s_{t-2} + \varepsilon_t\,, \qquad
  c_1 = \mathrm{Re}(z_1+z_2)\,,\quad c_2 = -\mathrm{Re}(z_1 z_2)\,,
\end{equation}
driven by i.i.d.\ Gaussian noise
$\varepsilon_t\sim\mathcal{N}(0,\sigma_\varepsilon^2)$.
This process can model underdamped
($\zeta<1$), critically damped ($\zeta=1$), and overdamped
($\zeta>1$) dynamics. The noise variance $\sigma_\varepsilon^2$ is
fixed by the unit-power constraint
\begin{equation}
  \sigma_\varepsilon^2 = G(\theta)^{-1}\,, \qquad
  G(\theta) = \frac{1-c_2}{(1+c_2)\,\bigl[(1-c_2)^2 - c_1^2\bigr]}\,,
\end{equation}
so that $\mathbb{E}_\theta[s_t^2]=1$ holds exactly for every
$\theta=(\zeta,f_0)$.
The two free parameters of the capacity
optimization are therefore the damping ratio $\zeta$ and the natural
frequency $f_0$ whereas the noise variance $\sigma_\varepsilon^2$
is determined by the power constraint.

% \paragraph{Bayesian optimization.}
% We maximize the mutual information over the two optimization
% parameters $\theta=(\zeta,f_0)$, the damping ratio and natural
% frequency, using Gaussian-process (Bayesian) optimization as
% implemented in \texttt{scikit-optimize}~\cite{skoptimize}.
% We search
% over $\zeta\in(10^{-2},10)$ and $f_0\in(0.5,25)\,\text{Hz}$.
% Each candidate $\theta$ is mapped to AR(2) coefficients $(c_1,c_2)$
% via the pole expressions above, from which the input distribution
% $P_\theta(s_{1:T})$ is fully determined. The optimizer is seeded
% with $8$ random evaluations of
% $I(\theta)$, which are used to fit a Gaussian-process surrogate
% model for $I(\theta)$.
% Subsequent candidate parameters are chosen by maximizing an
% acquisition function (expected improvement) over the surrogate,
% directing evaluations towards regions likely to improve on the
% current best. In total, we perform $100$ PWS evaluations. Because the MI
% estimator is itself stochastic, we model the observation noise with a
% fixed standard deviation of $0.3$~nats. The parameter configuration
% that achieves the largest estimated MI across all evaluations is
% reported as the channel-capacity estimate.

\paragraph{Grid search.}
We evaluate the mutual information on a fixed grid over the two optimization
parameters $\theta=(\zeta,f_0)$, the damping ratio and natural
frequency. The grid
comprises $8\times8=64$ points, with $\zeta$ logarithmically spaced over
$[0.1,10]$ and $f_0$ linearly spaced over $[0.5,20]\,\text{Hz}$.
Each candidate $\theta$ is mapped to AR(2) coefficients $(c_1,c_2)$ via the pole
expressions above, from which the input distribution
$P_\theta(s_{1:T})$ is fully
determined. At every grid point we draw $N=1000$ input trajectories of
$T=100$ time steps, i.e.\ a duration of $T\Delta t=\SI{1}{\second}$, from
$P_\theta$, generate the corresponding population responses by sampling from the
trained model, and estimate $I(S_{1:T};X_{1:T})$ by RR-PWS with
$M=4096$ Monte Carlo samples.
We report the information rate, obtained as the secant slope of
$I(S_{1:T};X_{1:T})$ between the reference time $\tau_0=\SI{0.5}{\second}$ and
the final duration $T\Delta t$ [\cref{eqn:SM-rate_secant}], which discards the
onset transient.
The largest rate on the grid,
$38.0\,\text{bit\,s}^{-1}$ at $\zeta=0.373$ and $f_0=14.43\,\text{Hz}$,
is reported as
the channel-capacity estimate in the main text.

\subsection{Computational Cost}
\label{sec:SM-ganglion-cost}

The computational costs of training the neural network model are
similar for all the ML-based estimators analysed in this study. In
comparison to the other ML-based estimators, ML-PWS does not only
require the training of a model, but also includes a marginalisation
step. While this marginalisation step gives the benefit of a rigorous
and reliable lower bound, as the benchmark examples illustrate, it
does entail a computational cost. To illustrate the costs of model
training and marginalisation, we quantify these costs for our most
realistic application, the neuronal system.

All model training and PWS estimates were computed on CPU. Training
the collective model (hidden size~256) for 40~epochs took $19\,\text{s}$ per
epoch, or 13~minutes in total, on 18~CPU cores. The 50 single-neuron
models (hidden size~20) are considerably cheaper to train: each took
2~minutes for 50~epochs. The cost of marginalisation is the same
order of magnitude as the cost of model training. Specifically, a
marginalisation step costs, on average, about 4~minutes for the
single-neuron model. These computations can be significantly
optimised using GPU acceleration.

\suppresstoc{\putbib[references,manuscriptNotes]}
\end{bibunit}

\end{document}